# Structural transitions related to order-disorder and thermal desorption of D atoms in $TbFe_2D_{4.2}$

V. Paul-Boncour[1]*, O. Isnard[2]

[1] *Université Paris-Est Créteil, CNRS, ICMPE, UMR7182, F-94320 Thiais, France*

[2] *Université Grenoble Alpes, Institut Néel, CNRS, BP166X, 38042 Grenoble Cédex 9, France*

* Corresponding author: valerie.paul-boncour@cnrs.fr

## Abstract

$TbFe_2D_{4.2}$ deuteride crystallizes in a monoclinic structure (*Pc* space group) with deuterium inserted into 13 [$Tb_2Fe_2$] and 5 [$TbFe_3$] tetrahedral interstitial sites. Its structural evolution versus temperature has been investigated by combining *in-situ* X-ray and neutron diffraction (XRD and NPD) with differential scanning calorimetry (DSC) experiments. Upon heating, the deuteride undergoes a reversible order-disorder transition from an ordered monoclinic structure to a disordered cubic structure between 320 and 380 K. Then, a multipeak thermal desorption occurs between 400 and 550 K, which can be explained by the transitions between different cubic deuterides separated by two-phase ranges. After controlled partial D desorption of $TbFe_2D_{4.2}$, the XRD patterns of several $TbFe_2D_x$ deuterides were measured *ex-situ* using synchrotron radiation at room temperature, revealing the formation of different phases with cubic ($x < 3$ and $x \geq 4.5$) or monoclinic structures ($3.6 \leq x \leq 3.8$, $x = 4.2(1)$) separated by two–phase ranges. A tetragonal superstructure was observed for a phase with $x \approx 2$ and the corresponding atomic positions of the metal atoms are reported. This work can explain previous results of the literature indicating the existence of cubic and/or rhombohedral hydrides depending on the hydrogenation conditions and the H or D content. The monoclinic structures reported here correspond to a slight distortion of the previous rhombohedral structures described by other authors.

## 1. Introduction

*R*$Fe_2$ Laves phase compounds (*R* = rare earth) have been widely studied for their magnetic properties, especially since the discovery of their giant magnetostrictive and magnetoelastic properties at room temperature, as reviewed by Clark [1, 2].

The hydrogen insertion in *R*$Fe_2$ compounds has also been investigated, revealing several phases with significantly varying structural properties depending on the quantity of inserted hydrogen atoms [3, 4]. These structural changes are related to lattice expansion and distortion caused by long-range hydrogen ordering in interstitial sites. Reversible order-to-disorder transitions have been observed upon heating, changing the structure of the hydrides from ordered to disordered cubic phases. [5-7]. The study of $R\mathrm{Fe_2H}_x$ compounds has shown that under moderate hydrogen pressure ($P$ < 1 MPa) the maximum capacity reaches 4.2-4.5 H/f.u.. The capacity can be raised to 5 H/f.u. by applying 1 GPa of hydrogen. $R\mathrm{Fe_2H_5}$ hydrides crystallize in an orthorhombic structure and remain stable at room temperature [8, 9]. A recent work has shown that $\mathrm{YFe_2}$ can absorb even up to 7 H/f.u. under several GPa of $\mathrm{H_2}$ in a diamond anvil cell, but this large capacity is not retained when the pressure is released to ambient pressure [10].

Several works on $\mathrm{TbFe_2H}_x$ hydrides have revealed that their crystal structures depend heavily on hydrogenation conditions, such as $\mathrm{H_2}$ pressure and temperature. Aoki et al. [11] studied the influence of the hydrogenation temperature on the hydride formation and the structural order under 1 MPa $\mathrm{H_2}$ pressure. A pressure-temperature phase diagram ( $P$ < 5 MPa and $T$ < 800 K) revealed the formation ranges of crystalline, amorphous, and decomposed hydrides [12]. At 445 K, the hydride crystallizes in a rhombohedral structure. Increasing the temperature to 620 K results in hydrogen-induced amorphization (HIA). Increasing the temperature further, up to 773 K, results in the precipitation of $\mathrm{TbH_x}$. Finally, at an even higher temperature, complete disproportionation occurs, resulting in mixture of α-Fe and $\mathrm{TbH_x}$. Transmission electron microscopy confirmed the amorphous state after heating at 620 K. These results were confirmed by Zajkov et al. [13] with similar results versus temperature. They found that hydrogenated crystalline $c$-$\mathrm{TbFe_2H_{4.2}}$ at 0.2 MPa retains the cubic structure of the parent compound, though with a cell volume expansion. The local environments of $c$-$\mathrm{TbFe_2H_{3.8}}$, amorphous $a$-$\mathrm{TbFe_2H_2}$, and $a$-$\mathrm{TbFe_2H_3}$ were investigated using the partial diffusion function (PDF), which was obtained from X-ray diffraction (XRD) and neutron powder diffraction (NPD) measurements [14]. Different local orders were observed between crystalline and amorphous hydrides. Berthier et al. [15] found that the $c$-$\mathrm{TbFe_2H}_x$ hydrides maintain a cubic structure for $x \leq 2.12$ and exhibit rhombohedral distortion at $x$ = 2.65 and 3.47 H/f.u. In cubic $AB_2$ Laves phases ($\mathrm{MgCu_2}$ type structure) there is three possible tetrahedral interstitial sites [$A_2B_2$], [$AB_3$] and [$B_4$] to insert H atoms. For the $c$-$\mathrm{TbFe_2H}_x$ hydrides, the [$\mathrm{Tb_2Fe_2}$] tetrahedral interstitial sites are filled at first as they are the biggest sites and present a better affinity with hydrogen related to the larger number of Tb atoms as near neighbors. The rhombohedral distortion was attributed to the additional filling of the [$\mathrm{TbFe_3}$] tetrahedral interstitial sites [15]. Because of their too small size as well as their less favorable atomic environment, the filling of [$B_4$] sites did not occur in such compounds. Kulshreshtha et al. [16] showed that a maximum content of 4.8 H/f.u. can be obtained at $P$ = 2.2 MPa from the pressure composition isotherm (PCI) at 298 K. They found that the crystalline hydrides keep a C15 cubic structure up to 4.8 H/f.u. with a progressive cell volume increase. Mushnikov et al. [17] found that the $c$-$TbFe_2H_x$ hydrides synthetized at different temperatures with a maximum H content up to 4 H/f.u. keep the cubic structure, except for $TbFe_2H_{3.56}$ which displays a rhombohedral distortion. In a previous study [18], we have shown from XRD patterns that $TbFe_2H_{4.2}$ and $TbFe_2D_{4.2}$ are both monoclinic ($C2/m$ space group) with a larger cell volume for the hydride compared to the deuteride.

This review of previous works raises several questions concerning the crystal structure of crystalline $TbFe_2$ hydrides. Are they cubic, rhombohedral or monoclinic depending on the range of H or D concentration?

To clarify discrepancies in the structures of $TbFe_2$ hydrides and deuterides observed in different publications and to solve the complexity of this system, it is necessary to systematically examine the relationship between nuclear structures and hydrogen or deuterium content. To this end, we report on the use of several complementary methods, including high-resolution powder X-ray and neutron diffraction experiments. Deuterides rather than hydrides were chosen as they are more suitable for neutron powder diffraction experiments.

First, we concentrate our efforts on fully characterizing the crystal structure of $TbFe_2D_{4.2}$ by XRD and NPD at room temperature. Additionally, we performed ex-situ XRD measurements at ambient temperature using high-resolution synchrotron radiation for given D contents. This allows us to present evidence of superstructures that had not been observed before for such deuterides. Then, we will present the evolution of $TbFe_2D_{4.2}$ upon heating under vacuum to determine structural changes versus temperature using in-situ XRD and NPD experiments, as well as differential scanning calorimetry (DSC). These experimental results will enable us to propose a structural phase diagram that can explain the discrepancies observed in the literature. This is the first time a systematic study of the crystal structure of $TbFe_2D_x$ system is provided.

These results will also be compared to those obtained for other $Y_{1-x}R_xFe_2$ hydrides and deuterides ($R$ = Pr, Gd, Tb, Er) [19-23], with the aim to understand the influence of the nature of the rare earth on the structure of the corresponding hydrides.

## 2. Experimental methods

The synthesis of $TbFe_2$ alloy was performed by induction melting of the pure elements followed by a 3-week annealing treatment at 1100 K under secondary vacuum. A small excess of Tb (5%) was added to compensate for any loss due to Tb oxidation during the heat treatment. The alloy composition was measured by electron probe microanalysis (EPMA) on a mirror-polished sample. The sample was found to be homogeneous and the average composition, measured at 62 different points, was $TbFe_{1.90(2)}$. This value shows a small deviation from the ideal 1:2 stoichiometry but agrees with the nominal composition of $TbFe_{1.9}$. Few inclusions of $Tb_2O_3$ were observed.

The $TbFe_2D_x$ deuteride was prepared using the Sieverts method with 7.5 g of alloy and deuterium gas. The final pressure was 2 bar, and the calculated D content was 4.2(1) D/f.u. The sample was quenched in liquid nitrogen, then slightly heated to room temperature in air to passivate the surface and prevent desorption of the deuterium, which can occur after several days. Deuterides with various D contents were obtained by controlling the desorption of this sample upon heating. One deuteride with a higher D content was prepared using the same high-pressure device as the $YFe_2H_5$ and $ErFe_2H_5$ compounds described in [8].

Differential Scanning Calorimetry (DSC) curves were measured using a Q100 instrument from Thermal Analysis (TA Instruments). Few mg of $TbFe_2D_{4.2}$ sample was placed in an aluminium container, which was pierced to evacuate the desorbed deuterium.

The XRD patterns were registered at room temperature on a Brucker D8 diffractometer using a Cu-Kα wavelength. The fluorescence of Fe was corrected by adjusting the energy window on the electronic of the detector. Some additional *in-situ* measurements were performed upon heating by using a Brucker D8 diffractometer equipped with a Vantec detector and Gobel X-ray mirrors. In this configuration, the fluorescence of the iron cannot be suppressed, but this provides good statistics that allow for a high signal-to-noise ratio in a short measurement time. Selected samples with different D concentrations were also measured using the synchrotron radiation beamline ID31 at ESRF (Grenoble, France) ($\lambda$ = 0.1652 Å) to obtain SR-XRD patterns with a better resolution than with the laboratory diffractometer and solve weak structural distortions.

The neutron powder diffraction experiments were performed in different neutron centers as described below. The NPD patterns have been recorded at 300 K on the high resolution 3T2 diffractometer at the Laboratoire Léon Brillouin (LLB, CEA, Saclay, France) with $\lambda$= 1.225 Å (15 ° < 2θ < 125 ° and a step of 0.05 °). The NPD patterns were also measured on high flux two axis diffractometer D1B ($\lambda$ = 2.52 Å) at the Institut Laue Langevin (ILL, Grenoble, France).

The measurements were done *in-situ* between 300-600K in an oven and the stainless-steel sample-holder connected to a vacuum pump. The pressure variation of the vacuum was registered to obtain a thermodesorption curve coupled to the NPD experiments.
All the XRD and NPD patterns were refined with the Rietveld method using the Fullprof code [24] to extract the structural parameters.

## 3. Results and discussion

### *3.1 Nuclear structure at 300 K*

The $TbFe_2$ sample is single phase and crystallizes in a rhombohedral structure with the following cell parameters: $a$ = 5.189(1) Å, $c$ = 12.757(1) Å and $V$= 297.47(2) Å$^3$ (given in the hexagonal triple unit cell). This structure corresponds to a distortion of the cubic cell along the (111) axis and attributed to a spontaneous magnetostriction as previously observed [25, 26].

The SR-XRD pattern of $TbFe_2D_{4.2}$ at 300 K was refined with a monoclinic structure described in $C2/m$ space group. The refined cell parameters are $a$ = 9.4423(2) Å, $b$ = 5.7494(5) Å, $c$= 5.5227(3) Å, $\beta$= 122.344(2)° and $V$ = 253.300(1) Å$^3$/f.u.. In addition, the diffraction pattern of the deuteride displays thin diffraction peaks without significant peak broadening compared to the initial alloys. This means that there is a long-range order without noticeable reduction of the crystallite size or induced microstrains.

The NPD pattern of $TbFe_2D_{4.2}$ measured at 300 K on the 3T2 diffractometer was refined in a monoclinic $Pc$ space group with a doubling of the cell parameter along the monoclinic ***b*** axis (Figure 1-a). This further lowering of crystal symmetry found in NPD pattern compared to XRD pattern is due to the ordering of D atoms, which cannot be observed by XRD. The new set of refined monoclinic cell parameters measured by NPD are $a$ = 5.5171(2) Å, $b$ =11.5061(5) Å, $c$= 9.4382(3) Å, $V$ = 506.19(3) Å$^3$ and $\beta$= 122.342(2)°. To determine the atomic positions and the occupation numbers of the D atoms a join refinement of the 3T2 pattern with one pattern measured at 300 K on D1B was performed (Figure 1a and 1b). The D atoms occupy only 18 interstitial sites (13 [$Tb_2Fe_2$] and 5 [$TbFe_3$] sites) among all possible sites (48 [$Tb_2Fe_2$] and 16 [$TbFe_3$] sites) generated by the lowering of crystal symmetry of the $MgCu_2$ cubic structure. This structure is close to that of $YFe_2D_{4.2}$, $Y_{1-y}Er_yFe_2$ an $Y_{0.9}Tb_{0.1}Fe_2D_{4.3}$ deuterides [19, 20, 27-29], with the same number of D atoms but with more [$TbFe_3$] interstitial sites (5 instead of 3). According to the Westlake criterion [30] these two types of interstitial sites have a radius of about 0.4 Å or larger, which is not the case for the [$Tb_4$] sites. All refined atomic positions on 2*a* Wyckoff site and occupancies factors are listed in Table 1. The total refined D content is 4.1(1) D/f.u., close to the value obtained by the Sievert method (4.2(1) D/f.u.). Compared to $YFe_2D_{4.2}$ a relative increase of the *a* (+0.15 %), *b* (+0.28 %), *c* (+0.25 %) and *V* (+0.73 %) cell parameters is observed due to the larger radius of Tb atoms compared to Y, whereas the monoclinic angle remains unchanged. Most of the distances between two D atoms are larger than 2 Å in agreement with Switendick criterion [31] and all occupied sites have a mean radius equal or larger than 0.39 Å (Table 1) in agreement with Westlake criterion [31]. However, one very short distance was found between D1 and D17 atom positions ($d$ = 1.27 Å). This can explain the low partial site occupation of these two sites ($N_{Occ}$ =0.5 for D1 and 0.60 for D17) as only one site over two can be occupied to avoid too close D atoms leading to strong electrostatic repulsion. Another short distance was found between the D2 and the D15 atoms with $d$=1.88 Å, slightly lower than 2 Å, and the occupation number of both sites is 0.90. This suggests that even if this distance is two short, this is an average value that can change from one cell to another. Interestingly, all these four D atoms have special common features, which are described below.

The D1 and D2 atoms are in [$Tb_2Fe_2$] interstitial sites whereas the D15 and D17 atoms are in [$TbFe_3$] interstitial sites. When plotting the environment of these four atoms, some common features are observed: they share some bonds with the same metal atoms. D1 and D17 are bonded to the same Fe6 and Tb2 atoms whereas D2 and D15 are both bonded to common Fe6 and Tb3 atoms. All the four D atoms are connected to the same Fe6 atom. D1, D2 and D17 are surrounding the same Tb2 atom and D1, D2 and D14 the same Tb3 atom. All these connections form a chain parallel to the ***c***-axis as illustrated in Figure 1c.

As detailed in Table S1 (supplementary materials), each Fe atom is surrounded by 3 to 5 D atoms, with Fe-D distances ranging from 1.54(6) to 2.52(4) Å, with average values between 1.6 and 2 Å for each Fe atom (supplementary material, Table S1). Each Tb atom, on the other hand, is surrounded by around 7±0.6 D atoms, with interatomic Tb-D distances ranging from 1.83(5) to 2.55(4) Å and average distances between 2.2 and 2.3 Å. The defined number of D neighbors around each metal atom can also explain that there is a long range ordering of the D atoms, which will occupy particular positions relative to the metal sublattice.

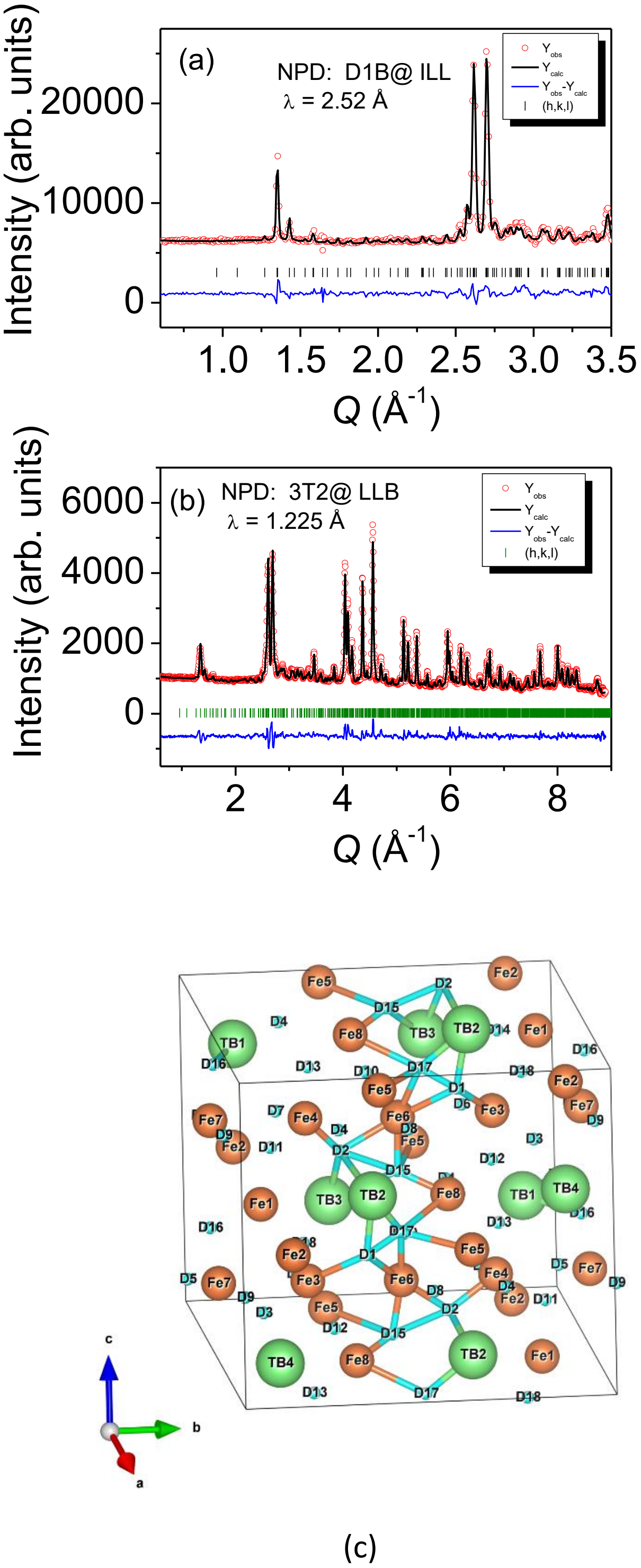

**Figure 1**: Neutron diffraction pattern of $TbFe_2D_{4.2}$ at 300 K measured with a) D1B diffractometer at ILL and b) 3T2 diffractometer at LLB (λ = 1.225 Å) refined by Rietveld method, (c) structure representation showing the connections of D1, D2, D15 and D17 atoms with first D, Tb and Fe neighbors.

**Table 1**: Atomic positions (*x,y,z*), occupation factors ($N_{occ}$), isotropic Debye Waller factor (*B*), type of D site and average radius of each site obtained by the join Rietveld refinement of $TbFe_2D_{4.2}$ NPD pattern (recorded at 300 K on 3T2 and D1B diffractometers) in *Pc* space group. The agreement factors are : $R_{Bragg}$ (3T2) : 7.68 %, $R_{Bragg}$ (D1B) : 10 %, Chi$^2$: 9.5. The $N_{Occ}$ parameters of D atoms were constrained to vary between 0 and 1.

| Name | *x* | *y* | *z* | $N_{Occ}$ | *B* (Å$^2$) | D site | $r_{site}$ (Å) |
|---|---|---|---|---|---|---|---|
| Tb1 | 0.1676 | 0.1138 | 0.8398 | 1 | 0.44(6) | | |
| Tb2 | 0.8898 | 0.3690 | 0.6018 | 1 | « « | | |
| Tb3 | 0.1341 | 0.3697 | 0.3364 | 1 | « « | | |
| Tb4 | 0.8943 | 0.1115 | 0.1011 | 1 | « « | | |
| Fe1 | 0.5231 | 0.1254 | 0.4648 | 1 | 0.54(3) | | |
| Fe2 | 0.0417 | 0.1291 | 0.4805 | 1 | « « | | |
| Fe3 | 0.5263 | 0.2474 | 0.2257 | 1 | « « | | |
| Fe4 | 0.5326 | 0.2477 | 0.7264 | 1 | « « | | |
| Fe5 | 0.9755 | 0.3795 | 0.9537 | 1 | « « | | |
| Fe6 | 0.5168 | 0.5005 | 0.2139 | 1 | « « | | |
| Fe7 | 0.5259 | 0.9929 | 0.7286 | 1 | « « | | |
| Fe8 | 0.5394 | 0.3732 | -0.0204 | 1 | « « | | |
| D1 | 0.7263 | 0.6249 | 0.8692 | 0.52(7) | 1.66(9) | $[Tb_2Fe_2]$ | 0.49(1) |
| D2 | 0.4212 | 0.6433 | 0.0842 | 0.91(3) | « « | $[Tb_2Fe_2]$ | 0.50(1) |
| D3 | 0.4980 | 0.1319 | 0.1218 | 1 | « « | $[Tb_2Fe_2]$ | 0.42(1) |
| D4 | 0.9025 | 0.2791 | 0.8125 | 1 | « « | $[Tb_2Fe_2]$ | 0.46(1) |
| D5 | 0.1661 | 0.0209 | 0.6230 | 1 | « « | $[Tb_2Fe_2]$ | 0.41(1) |
| D6 | 0.1510 | 0.2723 | 0.1234 | 1 | « « | $[Tb_2Fe_2]$ | 0.54(1) |
| D7 | 0.1510 | 0.7726 | 0.1227 | 0.82(9) | « « | $[Tb_2Fe_2]$ | 0.40(1) |
| D8 | 0.8894 | 0.4690 | 0.8018 | 1 | « « | $[Tb_2Fe_2]$ | 0.43(1) |
| D9 | 0.8676 | 0.9757 | 0.7974 | 1 | « « | $[Tb_2Fe_2]$ | 0.45(1) |
| D10 | 0.2858 | 0.4523 | 0.7776 | 1 | « « | $[Tb_2Fe_2]$ | 0.51(1) |
| D11 | 0.7408 | 0.7168 | 0.6478 | 0.82(8) | « « | $[TbFe_3]$ | 0.45(1) |
| D12 | 0.6001 | 0.2464 | 0.9071 | 0.96(2) | « « | $[Tb_2Fe_2]$ | 0.60(1) |
| D13 | 0.1535 | 0.8308 | 0.8465 | 0.94(9) | « « | $[TbFe_3]$ | 0.45(1) |
| D14 | 0.4990 | 0.9924 | 0.3911 | 0.82(8) | « « | $[Tb_2Fe_2]$ | 0.39(1) |
| D15 | 0.4222 | 0.5105 | 0.5167 | 0.90(8) | « « | $[TbFe_3]$ | 0.61(1) |
| D16 | 0.6530 | 0.1353 | 0.6807 | 0.97(7) | « « | $[Tb_2Fe_2]$ | 0.54(1) |
| D17 | 0.7197 | 0.4709 | 0.4312 | 0.60(8) | « « | $[TbFe_3]$ | 0.58(1) |
| D18 | 0.7532 | 0.2021 | 0.4186 | 0.94(7) | « « | $[TbFe_3]$ | 0.47(1) |
| total D content/f.u. | | | | 4.1(1) | | | |

*3.2 Structure of deuterides with various D contents at 300 K*

Following the neutron experiment, $TbFe_2D_{4.2}$ compound was heated under vacuum at different duration and temperatures between 400 and 473 K in order to desorb controlled quantity of deuterium. The desorbed D content was calculated with the application of the ideal gas law, knowing the pressure, the temperature and the desorption volume of the hydrogen device and sample holder. This yields deuterides with different D concentrations ranging from 1.3 to 4.0 D/f.u. All the SR-XRD patterns contain a mixture of crystalline phases related to a two- or three-phase range as well as small $Tb_2O_3$ peaks. The refined SR-XRD pattern of $TbFe_2D_{4.04(5)}$ at 300 K is presented as an example in Figure 2 and the result of the Rietveld refinement given in Table 2. This sample contains three $TbFe_2D_x$ phases with different weight percentages and 0.6 % of $Tb_2O_3$. The main $TbFe_2D_x$ phase (71.5 wt%) corresponds to the monoclinic $TbFe_2D_{4.2}$ phase which structure has been described before and with a cell volume by formula unit $V/Z$ = 63.396(3) Å$^3$/f.u.. The second $TbFe_2D_x$ phase (24.6(2) wt%) is also refined with a monoclinic structure ($C2/\underline{m}$ space group) with larger $a$ cell parameter, smaller $b$ and $c$ values and a smaller value of $V/Z$ = 62.042(2) Å$^3$/f.u..

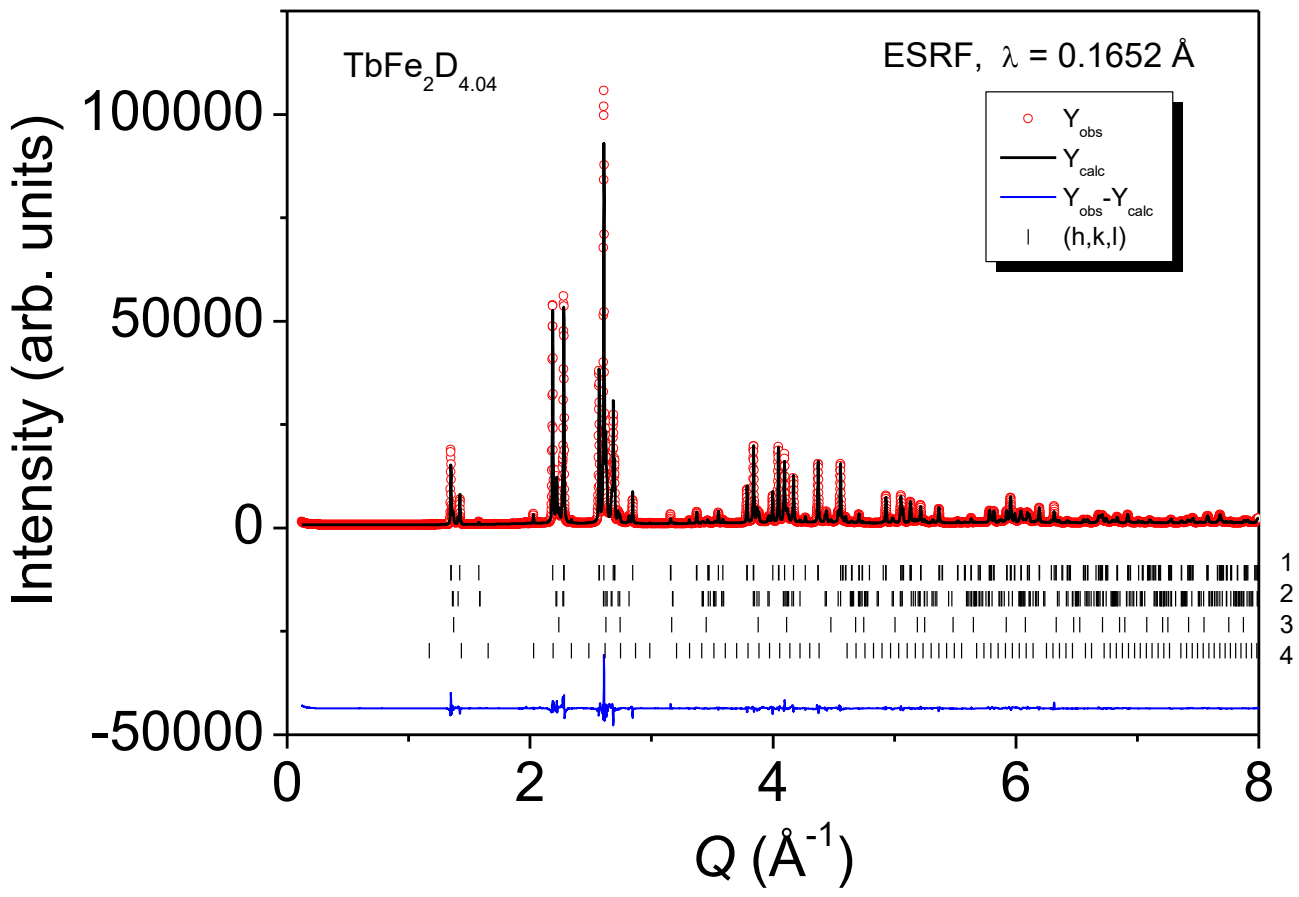

**Figure 2** : SR-XRD pattern of $TbFe_2D_{4.04}$ at 300 K measured on ID31 BL at ESRF (λ = 0.1652 Å) and refined using Rietveld method with three $TbFe_2D_x$ phases (referred to as phase 1 to phase 3) and $Tb_2O_3$ (phase 4). The agreement factors are $R_p$: 4.8 % and $R_{wp}$: 6.7 %. The four series of Bragg peaks correspond to the following phases with their space group, cell parameters and weight percentage reported in Table 2. The $R_{Bragg}$ factors for each phase are 3.4 % (phase 1), 5.9 % (phase 2), 6.4 % (phase 3) and 16.3 % (phase 4).

The cubic phase (3.3 (1) wt%) with $a$= 7.9405(2) has an intermediate cell volume by formula unit ($V$/Z = 62.583(3) Å$^3$/f.u.) between the two previous phases. These result agrees with those of $Y_{0.9}Gd_{0.1}Fe_2H_x$ compounds where the existence of one intermediate cubic phase between the two different monoclinic phases was observed [23]. However, it is difficult to isolate this cubic phase alone, as the equilibrium pressures between each monoclinic phase and this cubic phase are very close. This could explain why these three phases containing deuterium coexist.

Table 2 : Rietveld refinement results of the SR-XRD pattern of $TbFe_2D_{4.04}$ at 300 K presented in Fig.2. Space group, cell parameters and weight percentage of the three $TbFe_2D_x$ phases (1 to 3) and of $Tb_2O_3$ (4).

| Phase | Space group | $a$ (Å) | $b$ (Å) | $c$ (Å) | $\beta$ (°) | $V$ (Å$^3$) | Wt % |
|---|---|---|---|---|---|---|---|
| 1 | *C*2/*m* | 9.43754(9) | 5.75119(6) | 5.52375(8) | 122.518(9) | 252.809(5) | 71.5(3) |
| 2 | *C*2/*m* | 9.5247(1) | 5.67423(8) | 5.51831(8) | 123.683(1) | 248.169(6) | 24.6(2) |
| 3 | *Fd*-3*m* | 7.9405(2) | | | | 500.67(2) | 3.3 (1) |
| 4 | *I a* -3 | 10.7329(1) | | | | 1236.39(4) | 0.6(1) |

Five other $TbFe_2D_x$ compounds were prepared by D desorption ($x$= 1.3, 2.3, 3.0, 3.4 and 3.55 D/f.u.). In addition, a $TbFe_2D_{4.8}$ deuteride was synthetized under high deuterium pressure and it crystallizes in a cubic $MgCu_2$ type structure with $a$ = 7.987(1) Å.

The refined SR-XRD patterns for samples with D contents between 2.30 and 3.55 are presented in supplementary materials (Figures S1 to S4). The XRD patterns of the samples with $x$ = 1.3 and 4.8 were measured with a laboratory Bruker D8 diffractometer at 300 K with Cu $K_\alpha$ radiation. Each pattern contains at least a mixture of two $TbFe_2D_x$ phases with different cell parameters and few peaks due to $Tb_2O_3$ oxide. This indicates that we have a multiplateau behavior of the pressure composition isotherms as already observed for $YFe_2D_x$ and $Y_{0.9}Gd_{0.1}Fe_2H_x$ compounds [23, 32]. The results of the refinement with the phase percentages are reported in Table 2 (for $x$= 4.04) or in supplementary materials (Table S2). Table 3 reports the structure and the cell parameters of each individual $TbFe_2D_x$ phase sorted by increasing volume size. The volumes are expressed by formula units to compare compounds with different number of f.u. by cell. To estimate the deuterium content corresponding to each phase, we have assumed that the relation between the relative cell volume and the D content is similar to that for $YFe_2D_x$ compounds ($x$ = 0 to 5) and used a compilation of previous data [10]. The volume increase for $YFe_2D_x$ versus D content is not linear but follows a second-order polynomial equation:

$$V = f(x) = V_0 + x.V_1 - x^2 \cdot V_2 \quad (1)$$

With $V_0$ = 49.83 Å$^3$/f.u., $V_1$ = 4.22 Å$^3$/ (D atom), $V_2$ = 0.24 Å/ (D atom)$^2$

The $x$ value for each $TbFe_2D_x$ phase was calculated using this equation and the difference in volume $\Delta V = V - V_0$, with $V_0$ = 49.578 Å$^3$ (Table 3).

**Table 3**: Cell parameters at 300 K of different $TbFe_2D_x$ deuterides obtained by D desorption from $TbFe_2D_{4.2}$.* and **: the XRD patterns were only measured with the D8 Bruker diffractometer ($\lambda$ =Cu $K_\alpha$). ** the deuteride was synthetized under high D pressure (0.8 GPa). The $x$ content is estimated using the same cell volume expansion than for $YFe_2D_x$ compounds.

| Structure | Space group | *a* (Å) | *b* (Å) | *c*(Å) | *β* (°) | *V*/Z (Å$^3$/f.u.) | *x* (D/f.u.) |
|---|---|---|---|---|---|---|---|
| Cubic | *R*-3*m* | 5.189(1) | | 12.757(1) | | 49.578(1) | 0.00 |
| Cubic* | *Fd*-3*m* | 7.367(1) | | | | 49.970(1) | 0.09 |
| Cubic* | *Fd*-3*m* | 7.574(1) | | | | 54.310(2) | 1.20 |
| Tetragonal | *I*-4 | 12.1851(2) | | 23.1409(7) | | 57.271(2) | 2.05 |
| Cubic | *Fd*-3*m* | 7.7555(1) | | | | 58.292(2) | 2.39 |
| Cubic | *Fd*-3*m* | 7.8355(1) | | | | 60.133(2) | 3.02 |
| Cubic | *Fd*-3*m* | 7.8417(1) | | | | 60.275(2) | 3.07 |
| Cubic | *Fd*-3*m* | 7.8773(1) | | | | 61.100(2) | 3.38 |
| Monoclinic | *C*2/*m* | 9.5213(1) | 5.6662(1) | 5.5164(7) | 123.7170(7) | 61.886(1) | 3.69 |
| Monoclinic | *C*2/*m* | 9.5247(1) | 5.67423(8) | 5.51831(8) | 123.684(1) | 62.042(2) | 3.76 |
| Cubic | *Fd*-3*m* | 7.9405(2) | | | | 62.583(3) | 3.98 |
| Monoclinic | *C*2/*m* | 9.4375(1) | 5.7512 (1) | 5.5238(1) | 122.518(6) | 63.202(1) | 4.28 |
| Monoclinic | *C*2/*m* | 9.442(1) | 5.755(1) | 5.520(1) | 122.29(1) | 63.396(3) | 4.35 |
| Cubic** | *Fd*-3*m* | 7.987(1) | | | | 63.695(2) | 4.49 |

Due to the high resolution and good signal/noise ratio of the synchrotron experiments, it was possible to observe several additional tiny peaks in the pattern of the sample $TbFe_2D_{2.30}$. Most of them belong to $TbFe_2D_{2.05}$ phase and can be indexed in a tetragonal supercell isostructural to that of $YFe_2D_{1.9}$ [5]. The structure of this phase was described in the *I*-4 space group with:

$a_{tetra}$ $=a_{cub}.\sqrt{5}/2$ and $c_{tetra}$ $=3.c_{cub}$ for a simple cubic phase with $a_{cub}$ = 7.7092(1). The refined cell parameters for this supertructure are $a_{tetra}$ = 12.1851(2) Å and $c_{tetra}$ = 23.1409(7) Å. The lowering of crystal symmetry from the cubic structure (*Fd*-3*m*) to this tetragonal cell (*I*-4) yields 10 Tb and 15 Fe Wyckoff sites (Table 4). A Rietveld refinement of the pattern taking into account this superstructure was performed using the results obtained by J. Ropka in her Phd thesis for $YFe_2D_{1.9}$ [33]. A zoom of the refined SR-XRD pattern of $TbFe_2D_{2.30}$ comparing the refinement with simple cubic structure (*F*-43*m*) or the tetragonal superstructure (*I*-4) is reported in Figure 3. In both refinement the contribution of cubic $YFe_2D_{2.39}$ and cubic $Tb_2O_3$ was considered.

**Table 4**: Atomic positions of the $TbFe_2D_{2.05}$ phase described in the *I*-4 space group with *a* = 12.1864(2) Å and *c* = 23.1368(7) Å obtained by the refinement of the SR-XRD pattern recorded for $TbFe_2D_{2.3}$ sample.

| Name | site occ | *x* | *y* | *z* | *B* (Å$^2$) |
|---|---|---|---|---|---|
| Tb1 | 2*a* | 0 | 0 | 0 | 0.501(62) |
| Tb2 | 4*e* | 0 | 0 | 0.333(3) | $B_{Tb}$ |
| Tb3 | 2*c* | 0 | 0.5 | 0.25 | $B_{Tb}$ |
| Tb4 | 4*f* | 0 | 0.5 | 0.580(4) | $B_{Tb}$ |
| Tb5 | 8$g_1$ | 0.574(5) | 0.709(5) | 0.919(2) | $B_{Tb}$ |
| Tb6 | 8$g_2$ | 0.600(5) | 0.715(4) | 0.586(3) | $B_{Tb}$ |
| Tb7 | 8$g_3$ | 0.889(4) | 0.292(5) | 0.163(2) | $B_{Tb}$ |
| Tb8 | 8$g_4$ | 0.605(4) | 0.708(5) | 0.241(2) | $B_{Tb}$ |
| Tb9 | 8$g_5$ | 0.909(3) | 0.297(4) | 0.832(2) | $B_{Tb}$ |
| Tb10 | 8$g_6$ | 0.892(5) | 0.301(5) | 0.508(2) | $B_{Tb}$ |
| Fe1 | 8$g_7$ | 0.966(6) | 0.408(6) | 0.041(3) | 0.974(63) |
| Fe2 | 8$g_8$ | 0.779(6) | 0.980(5) | 0.720(3) | $B_{Fe1}$ |
| Fe3 | 8$g_9$ | 0.935(6) | 0.374(6) | 0.708(3) | $B_{Fe1}$ |
| Fe4 | 8$g_{10}$ | 0.958(7) | 0.893(7) | 0.544(3) | $B_{Fe1}$ |
| Fe5 | 8$g_{11}$ | 0.846(6) | 0.700(6) | 0.544(3) | $B_{Fe1}$ |
| Fe6 | 8$g_{12}$ | 0.751(6) | 0.979(5) | 0.368(3) | $B_{Fe1}$ |
| Fe7 | 8$g_{13}$ | 0.869(6) | 0.217(7) | 0.709(3) | $B_{Fe1}$ |
| Fe8 | 8$g_{14}$ | 0.841(5) | 0.198(5) | 0.382(3) | $B_{Fe1}$ |
| Fe9 | 8$g_{15}$ | 0.949(6) | 0.900(6) | 0.210(4) | $B_{Fe1}$ |
| Fe10 | 8$g_{16}$ | 0.954(6) | 0.912(6) | 0.872(3) | $B_{Fe1}$ |
| Fe11 | 8$g_{17}$ | 0.849(5) | 0.218(4) | 0.037(3) | $B_{Fe1}$ |
| Fe12 | 8$g_{18}$ | 0.834(5) | 0.692(5) | 0.873(3) | $B_{Fe1}$ |
| Fe13 | 8$g_{19}$ | 0.849(6) | 0.704(6) | 0.216(3) | $B_{Fe1}$ |
| Fe14 | 8$g_{20}$ | 0.743(6) | 0.999(5) | 0.041(3) | $B_{Fe1}$ |
| Fe15 | 8$g_{21}$ | 0.966(6) | 0.398(5) | 0.374(3) | $B_{Fe1}$ |

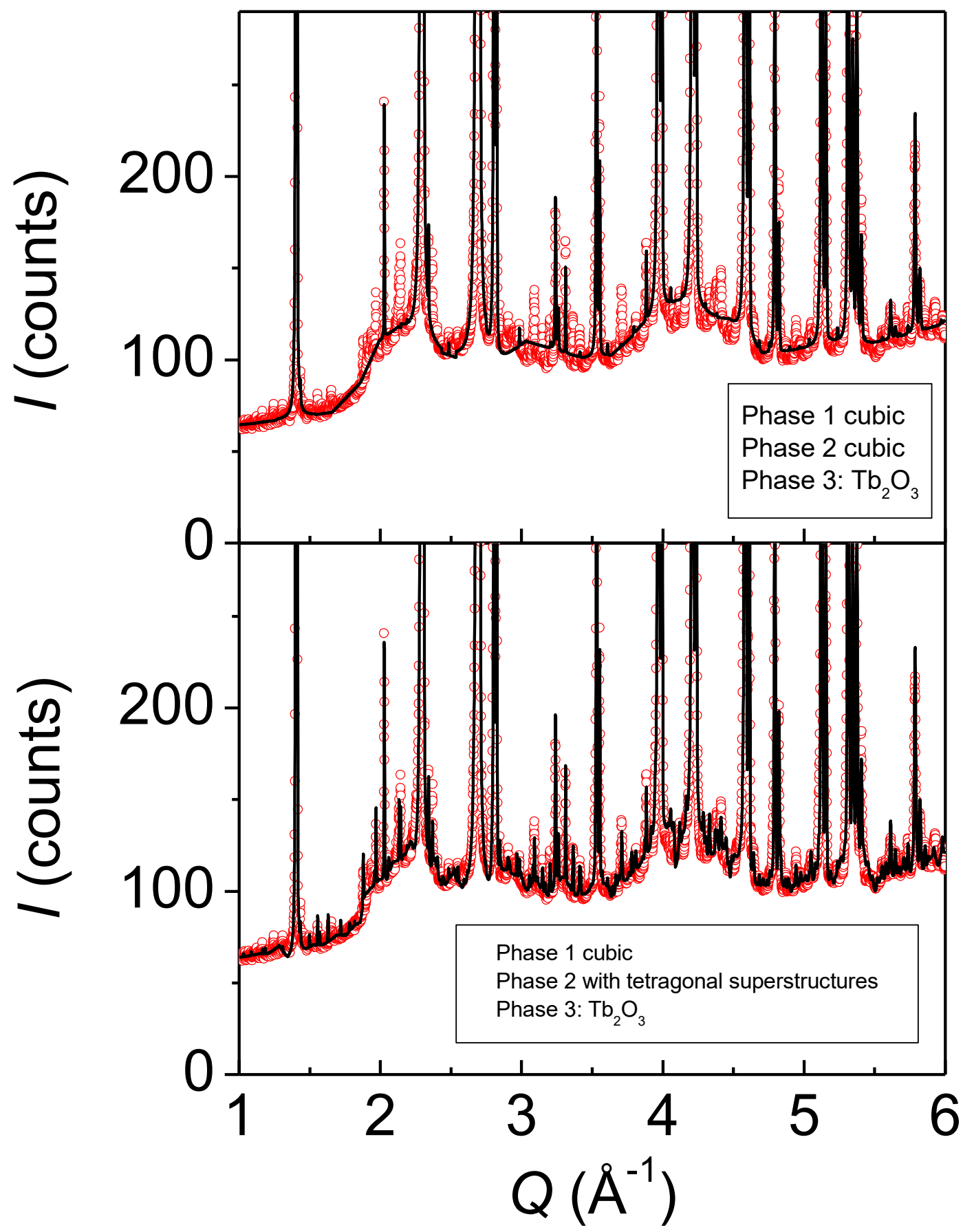

**Figure 3** : Refined SR-XRD pattern (λ = 0.1652 Å) of $TbFe_2D_{2.30}$. Zoom of the pattern showing: top) Rietveld refinement with two $TbFe_2D_x$ cubic phases ($x$ = 2.05 and 2.39) and $Tb_2O_3$, bottom) Rietveld refinement with one $TbFe_2D_{2.39}$ cubic phase, tetragonal $TbFe_2D_{2.05}$ phase with superstructure and $Tb_2O_3$.

To summarize, most of the deuterides with $x < 3.5$ crystallize in a cubic $MgCu_2$ type structure with an increase of the cell parameter versus D content, but for $TbFe_2D_{2.05}$ superstructure peaks refined in a tetragonal cell are observed. In addition, two different types of $TbFe_2D_x$ monoclinic compounds with $3.7 \leq x \leq 3.8$ D/f.u. and $x = 4.2\pm0.2$ are observed. These monoclinic structures can be obtained by the lowering of crystal symmetry from cubic *Fd*-3*m* to monoclinic *C*2/*m* space group through an intermediate rhombohedral *R*-3*m* space group. The refinement in *Pc* space group can only be observed by neutron diffraction as it is related to the ordering of D atoms.

To estimate the amplitude of the monoclinic distortion, it is possible to calculate the monoclinic $a$, $b$, $c$ and $\beta$ cell parameters considering only the lowering of crystal symmetry without any distortion and starting from an equivalent cubic cell parameter with the same cell volume by f.u.. The two monoclinic phases display different monoclinic $\beta$ angles 123.83(1)-123.68(1)° for $x \approx 3.5$ and 122.29(1)° for $x = 4.2$, i.e. smaller than the calculated non-distorted value $\beta$ = 125.26°. This indicates that a lower value of $\beta$ means a larger distortion. In addition, for $x \approx$ 3.76: the calculated non distorted parameters are $a$ = 9.6964 Å, $b$ = $c$ = 5.5982 Å, which compared to the refined monoclinic cell parameters $a$ = 9.5247(1) Å, $b$ = 5.6742(1) Å, $c$= 5.5183(1) Å indicates a contraction in the ($a$, $c$) plane and an expansion along the equivalent ***b*** axis. A similar observation can be made for $x$ = 4.2(1), as the calculated non-distorted values in *C*2/*m* space group: $a$ = 9.7670 Å, $b$ = $c$= 5.639 Å show also a reduction of $a$, $c$ and $\beta$ parameters and an increase of $b$. In addition, in the range $3.69 < x < 3.76$ the three set of parameters ( Table 3) are related to a range of solubility of the first monoclinic phase as they are not coexisting with each other but are in equilibrium with either a cubic and/or a monoclinic phase. As the D content increases, we observe an increase of the $a$, $b$ and $c$ cell parameters and a slight decrease of the monoclinic angle. Similar results were observed in the phase diagram of $Y_{0.9}Gd_{0.1}Fe_2H_x$ [23].

The existence of several phases with different structure types (cubic, tetragonal or monoclinic) forming within a narrow range of hydrogen or deuterium concentration can explain why different authors have obtained different results, depending on their hydrogenation conditions. In several work a rhombohedral distortion was described for large H content. In the present study, we have observed, thanks to the high instrumental resolution accessible with synchrotron radiation, that these “rhombohedral phases” should be in fact monoclinic [27] and exists for the concentration $3.69 < x < 3.76$ and 4.2 H(D)/f.u. For larger H/D content ($x$ = 4.5(1)) a cubic phase is observed. In addition, the present results demonstrate a strong similarity between the $TbFe_2D_x$ and the $YFe_2D_x$ structural phase diagrams at room temperature. Several different type of structures are present due to the long-range ordering of D atoms. For example, the tetragonal superstructure of $RFe_2D_{2(1)}$ deuteride and the monoclinic distortions for $3.6 \leq x \leq 4.2$ are found in both system. However, the tetragonal $YFe_2D_{1.3}$ and cubic $YFe_2D_{1.75}$ superstructures were not observed among $TbFe_2D_x$ deuterides. In addition, the orthorhombic $YFe_2D_5$ phase was not found under high D pressure for this Tb compound. This would require a more detailed analysis for low and high D content to conclude whether they exist also for Tb containing compound.

*3.3 Order-disorder transition and thermal desorption*

a) DSC measurements

The differential scanning calorimetry (DSC) curve, measured at a rate of 5 K/min, shows two peaks with maxima at 320 and 329 K during heating and at 314 and 326 K during cooling (Figure 4a). These peaks are exothermic upon heating and endothermic upon cooling, with an estimated heat flow of 2-3 J/g for each peak. They correspond to two reversible first-order structural transitions. This interpretation is supported by earlier results on $R$Fe$_2$H(D)$_{4.2}$ isotype compounds, which have experienced an order-disorder (OD) transition of H(D) in a similar temperature range. This hypothesis will be discussed later. At higher temperature, the DSC curve displays two new broad exothermic peaks at 466 K and 515 K (Figure 4b). The first peak displays also two small shoulders at 445 and 466 K. The total heat flow deduced from the surface integration is around 320 J/g and can be attributed to deuterium desorption. Upon cooling a small peak with a maximum at 693 K (0.5 J/g) is found, whereas the other peaks have disappeared. This peak corresponds to the Curie temperature of TbFe$_2$, which was measured at 694 K in [34], meaning that full D desorption has been achieved.

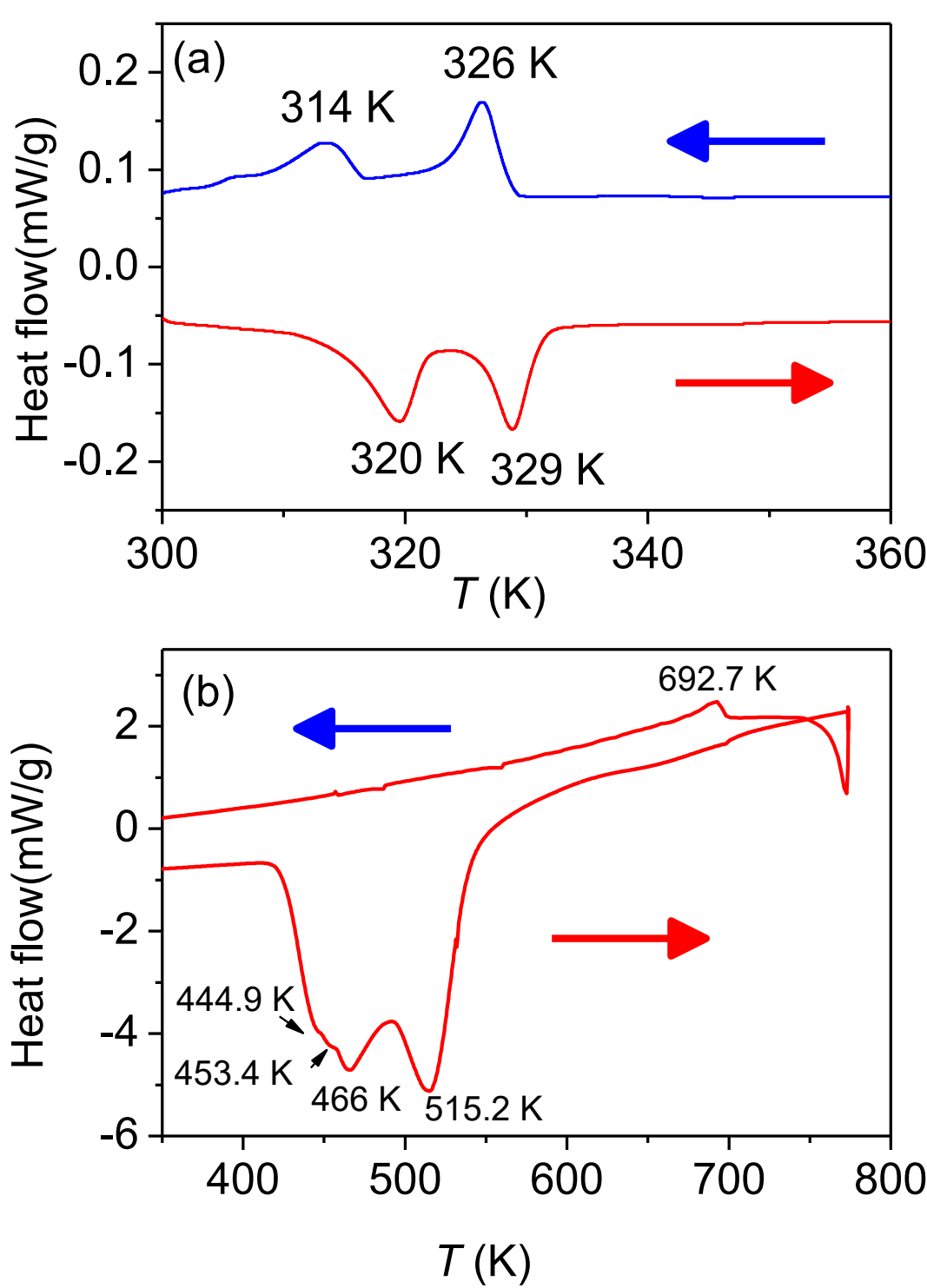

**Figure 4**: Heat flow as a function of temperature for the deuteride TbFe$_2$D$_{4.2}$ (5 K/min), (a) order-disorder transitions, (b) thermal desorption.

b) Thermal desorption curve

Pressure variation was recorded during the *in-situ* neutron diffraction experiment on D1B instrument to determine the quantity of desorbed deuterium and the thermal desorption rate (Figure 5). The multipeak behavior is similar to that observed for $YFe_2D_{4.2}$ [35] and must probably reflect the similarities of the interstitial sites occupy by D atoms in both systems. Desorption begins around 400 K and ends near 550 K, which aligns with the DSC curves. Two groups of peaks are observed, with several maxima identified at 431, 452, 507, and 525 K.

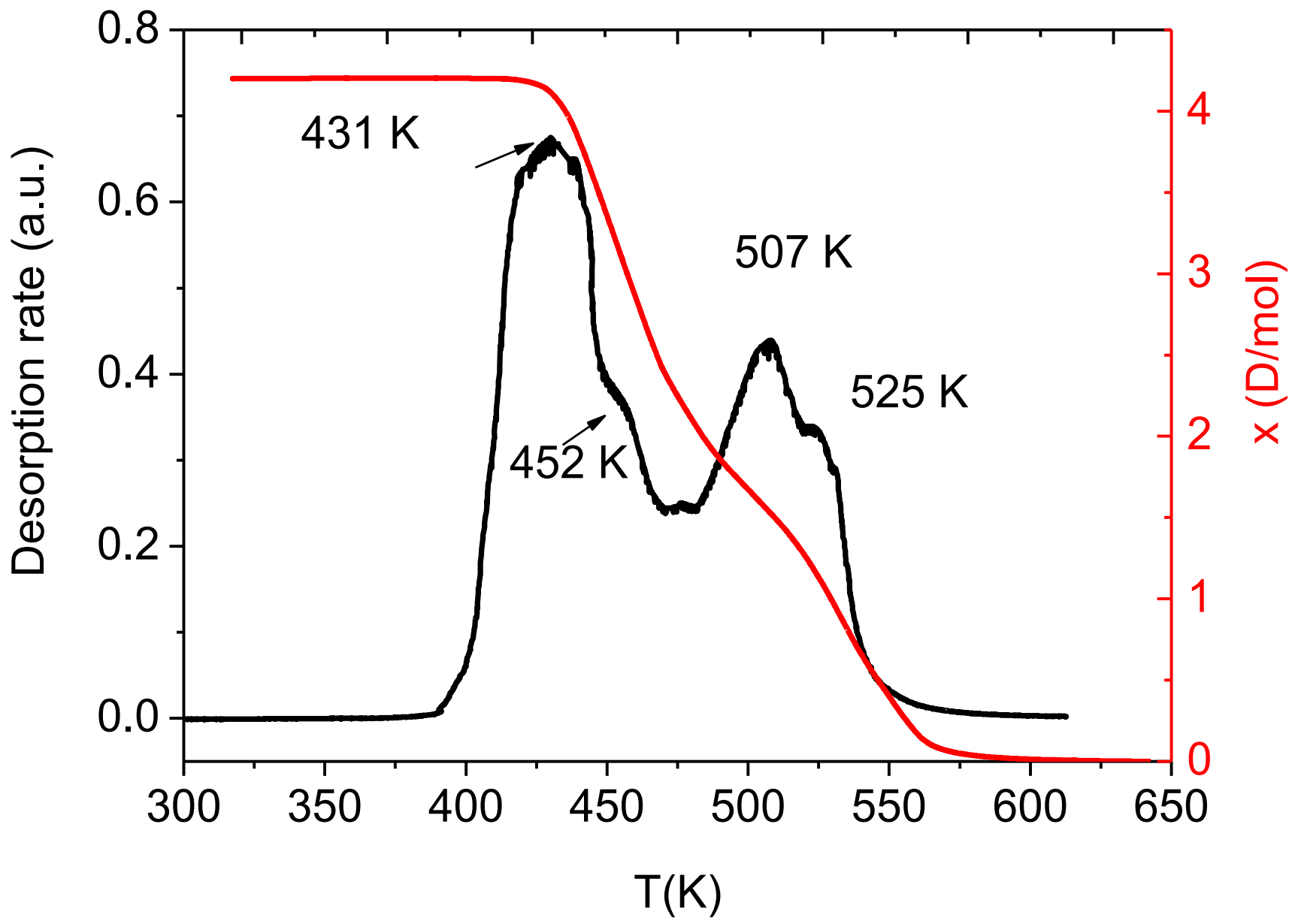

**Figure 5**: Thermal desorption rate registered during the neutron diffraction experiments on 3T2 diffractometer for $TbFe_2D_{4.2}$. In red (left axis) the calculated D content.

As for $YFe_2D_{4.2}$ isotype compound, each peak can be attributed to the transition between two deuterides with different D content and structures formed upon D desorption. This agrees with the observation of the existence of different phases described in part 3.2.

**c**- X-ray and neutron powder diffraction

The patterns were measured using *in-situ* X-ray and neutron diffraction when heated to observe the structural evolution versus temperature. A 3D plot of the NPD measurements on D1B is presented in Figure 6. First, the reversible order-disorder transition is presented and discussed, followed by the thermal desorption results.

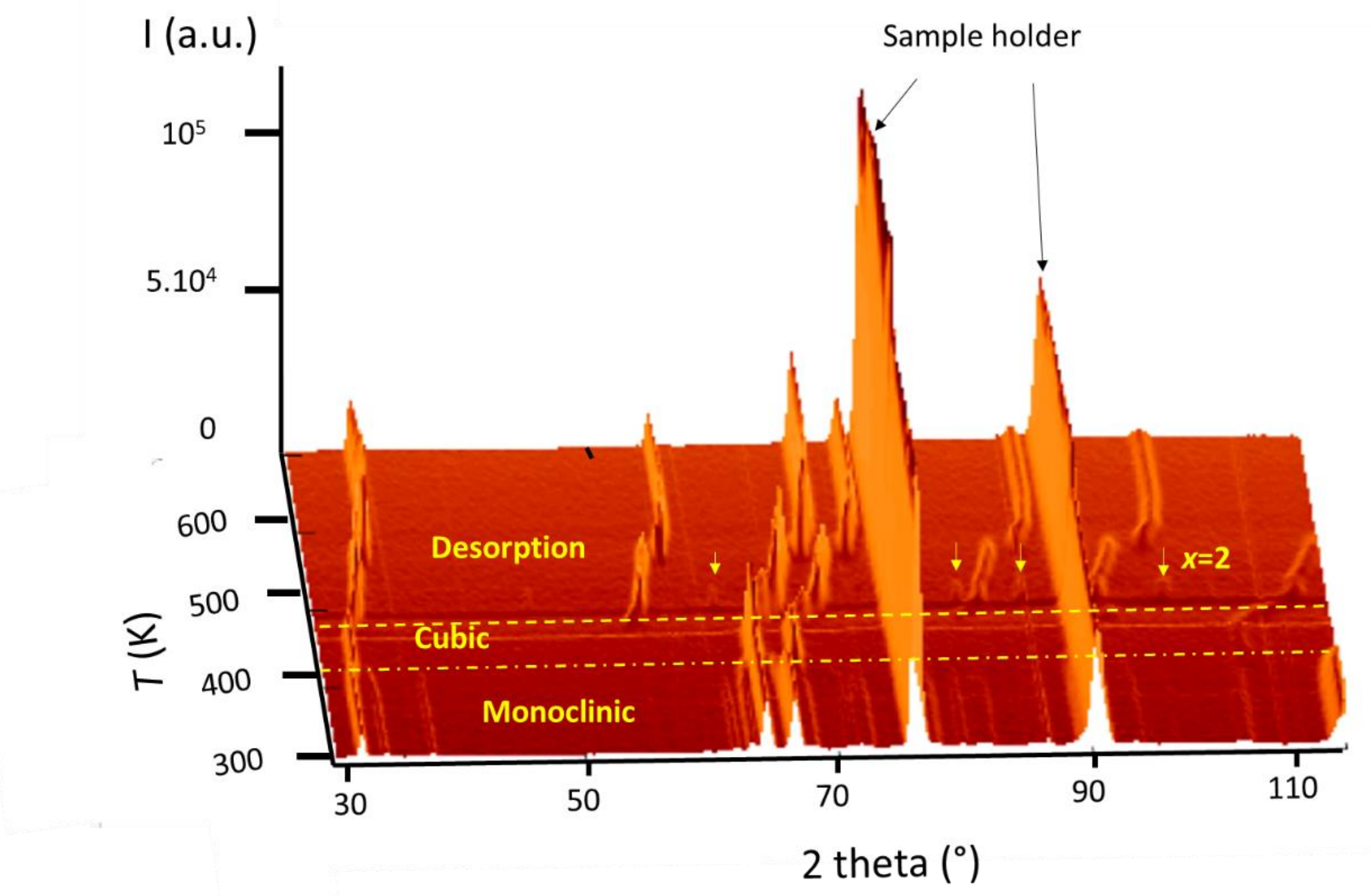

**Figure 6:** 3D plot of the evolution of the NPD patterns of $TbFe_2D_{4.2}$ measured on D1B upon heating from 295 K to 620 K. (λ = 2.52 Å). The structures of the phases are indicated in the plot as well as the desorption range. The small arrow indicated the superstructure peaks of $TbFe_2D_2$.

*Order-disorder transition*

Figure 7 displays the XRD and NPD patterns measured at selected temperatures. The variation of the peak positions in a 2D contour plot (blue= low intensity and red =high intensity) is presented in Figure S5 in supplementary materials. The transition from a monoclinic phase to a cubic phase is observed in both the XRD and NPD patterns between room temperature and 380 K [7, 27].

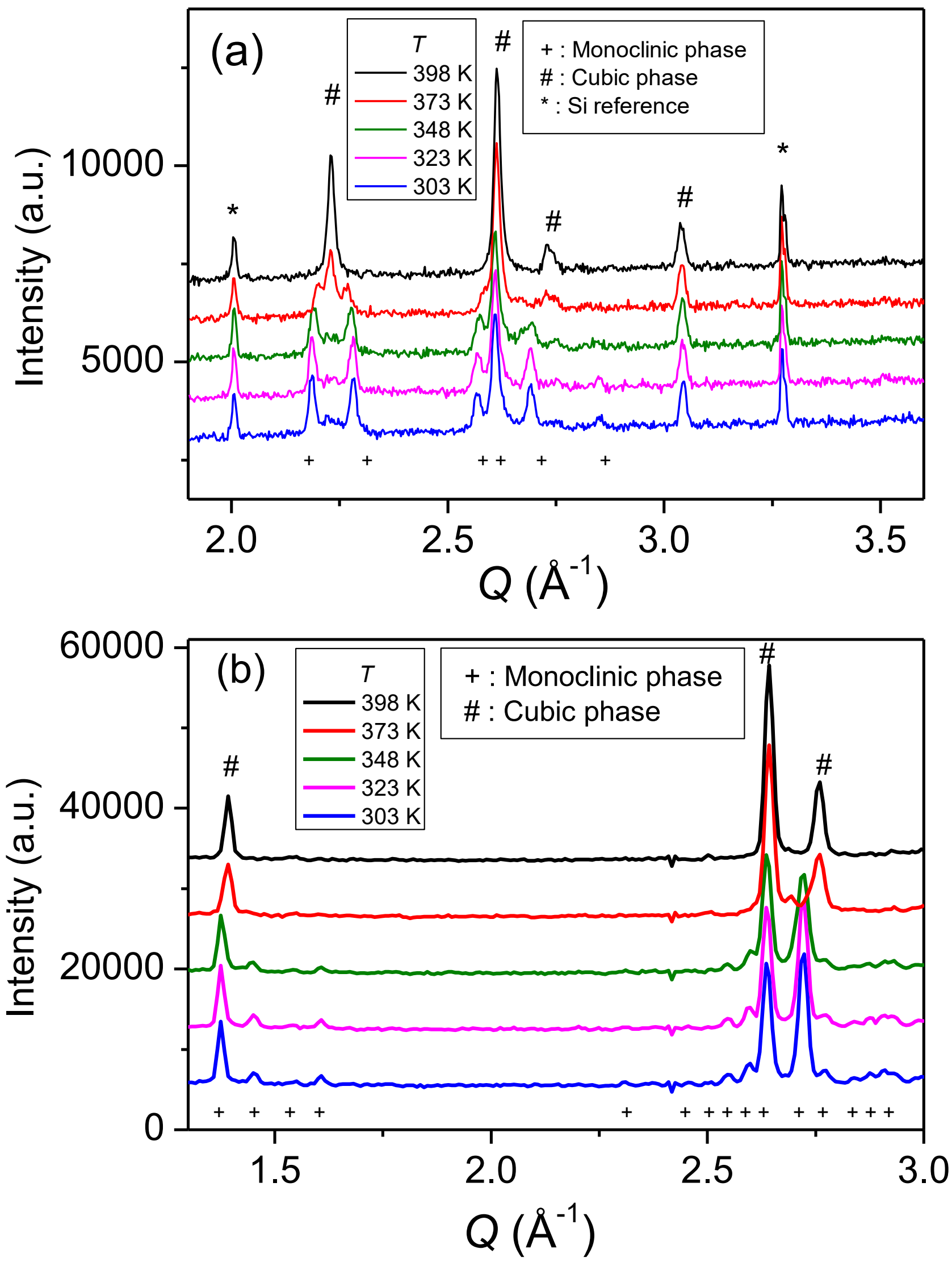

**Figure 7**: (a) XRD (λ of Cu $K_{\alpha}$, 1.5418 Å) and (b) NPD patterns measured on D1B ( λ =.52 Å) of $TbFe_2D_{4.2}$ at indicated temperatures between 300 and 400 K showing the monoclinic-cubic transformation. In the case of XRD patterns, a Si powder was added as reference to the sample to calibrate the temperature.

The low instrumental resolution of the laboratory X-ray diffractometer used to register the *in-situ* XRD pattern of $TbFe_2D_{4.2}$ deuteride upon heating prevented the refinement of the patterns considering monoclinic distortion. Thus, the patterns at low temperatures were refined using a mean rhombohedral structure (*R*-3*m* space group). Figure 8 displays the evolution of the cell parameters and the weight percentages of the rhombohedral and cubic phases in $TbFe_2D_{4.2}$ versus temperature.

The presence of 20 wt% of cubic phase at 300 K indicates that the sample was already on the equilibrium plateau pressure of the Pressure-Composition Isotherm between the cubic and monoclinic phase. The structural transition occurs through a two-phase range between 360 and 400 K, indicating that this transition is of the first order, in agreement with DSC results. Upon heating above 340 K, the rhombohedral distortion progressively decreases with a reduction of the *a* cell parameter and an increase in the *c* parameter. The refinement of the NPD patterns confirms a monoclinic - cubic transition with a two-phase range between $T_{OD}$ = 350 and 380 K. Even though the NPD-instrumental resolution is smaller than that by XRD, the presence of superstructure peaks induced by D-ordering allows one to distinguish the monoclinic structure from the rhombohedral one on the diffraction patterns.

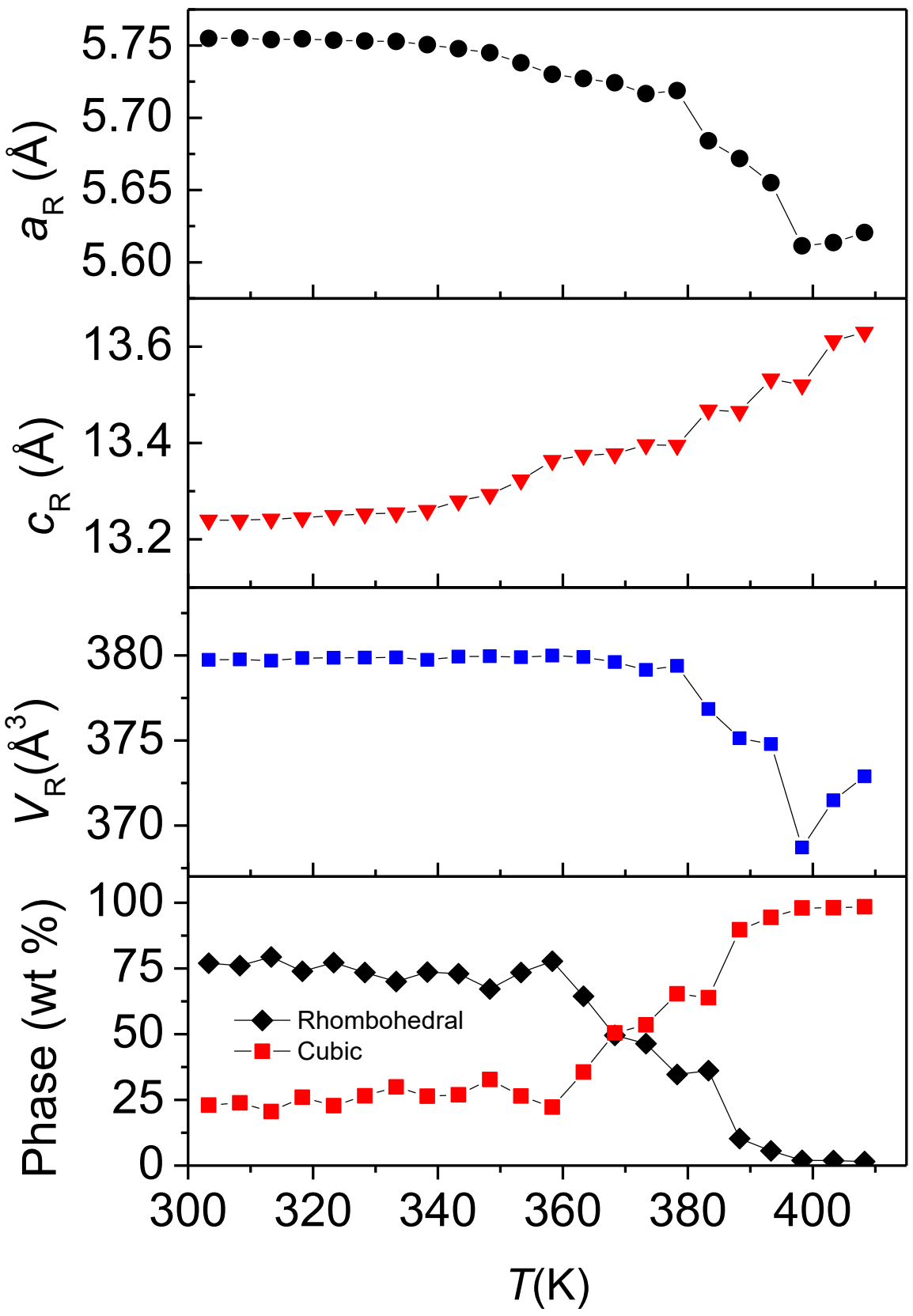

**Figure 8**: Temperature evolution of the rhombohedral cell parameters and the weight percentages of rhombohedral and cubic phases obtained from the Rietveld refinement XRD patterns of $TbFe_2D_{4.2}$ compound upon heating.

Analysis of the NPD pattern refinement shows that the monoclinic distortion is reduced with an expansion of *a*, *c* and $\beta$ cell parameters and a contraction of *b* cell parameter (Figure 9).

However, when the cubic phase appears at 360 K, the $c$ parameter and the cell volume of the monoclinic phase decreases. Taking into account the lowering of crystal symmetry from cubic (SG *Fd*-3*m*, # 227) to monoclinic structure (SG *P*c, # 7), for $a_{\text{cubic}}$ =7.96 Å, we obtain $a_{\text{mono}}$ = 5.620 Å, $b_{\text{mono}}$ = 11.240 Å, $c_{\text{mono}}$ = 9.734 Å and $\beta$= 125.264 °. Compared to the cell parameters obtained by NPD patterns refinement at 380 K, we can conclude that also reduced, the monoclinic distortion is still present. The decrease of $c_{\text{mono}}$ and $V_{\text{mono}}$ above 360 K, suggests that a small deuterium desorption of the monoclinic phase occurs, with a possible diffusion towards the cubic phase. Note that XRD data also observed a decrease in cell volume for the rhombohedral phase (Figure 8).

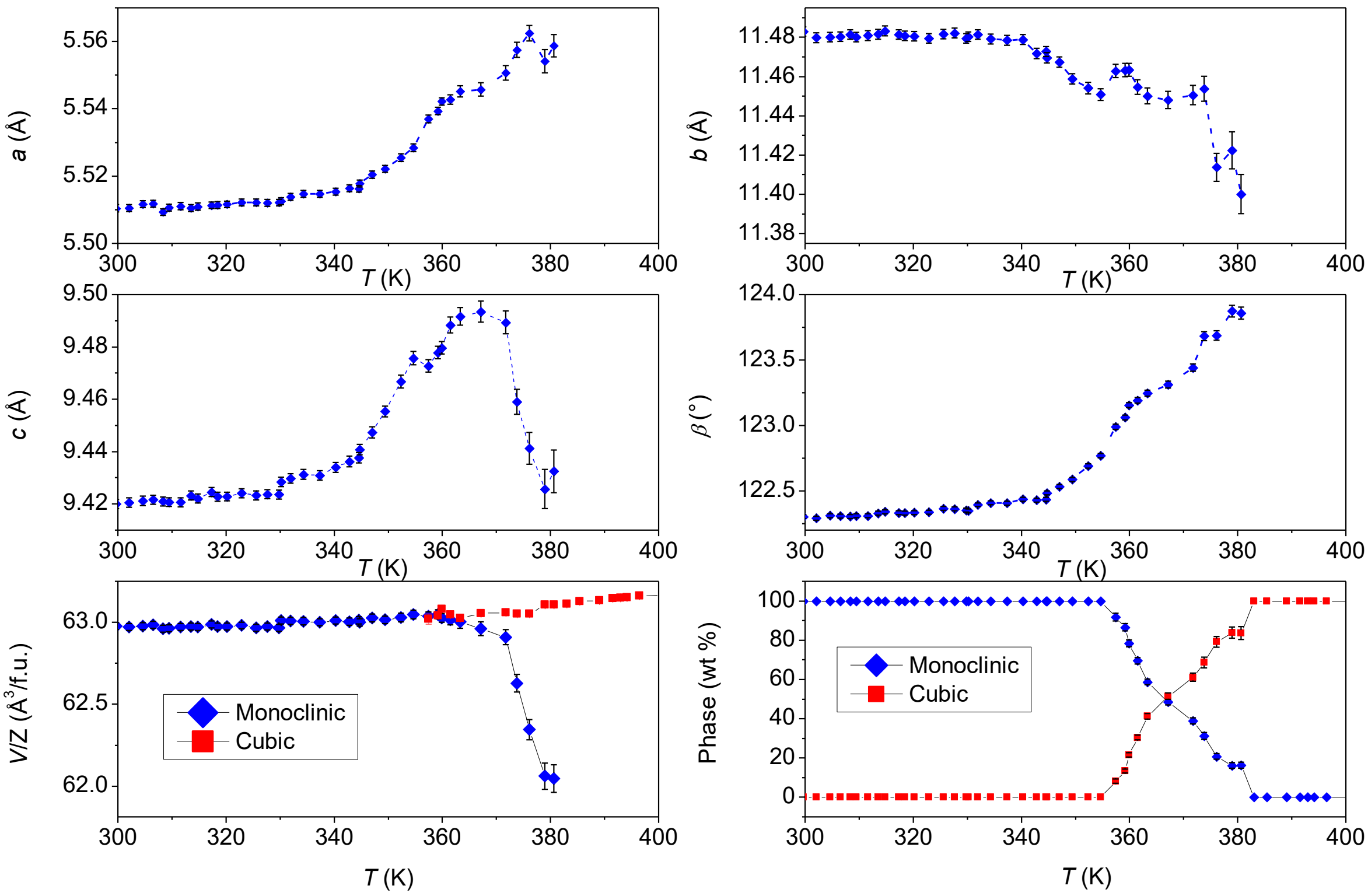

**Figure 9:** Thermal variation of the monoclinic and cubic cell parameters of $TbFe_2D_{4.2}$ in the reversible Order-Disorder range as refined from the NPD pattern recorded on D1B diffractometer upon heating from 300 to 400 K.

In the monoclinic structure the D atoms are inserted in 13 [$Tb_2Fe_2$] sites and 5 [$TbFe_3$] sites. Upon heating the transformation toward a cubic structure in *Fd*-3*m* space group reduces the number of sites to 1 [$TbFe_2$] site and 1 [$TbFe_3$] site with a statistical D occupation. The refinement of the NPD pattern at 404 K with one single cubic phase shows that 3.64(8) D/f.u.

atoms are in the [$Tb_2Fe_2$] site and 0.55(8) D/f.u. in [$TbFe_3$], corresponding to a total of 4.2 D/f.u. in agreement with the initial D content. Previous works on isostructural $YFe_2D_{4.2}$ has shown that even if the long range D order is lost, the local order is maintained at the scale of one unit cell ($d$ =8 Å) [7].

The transition temperatures measured by XRD and NPD are significantly higher than those measured by DSC. This discrepancy may be due to the different heating rates, as well as the inertia required to heat several grams of the sample in a silica tube within the NPD oven versus 10 mg of powder in an aluminum container within the DSC.

*Deuterium desorption*

The changes of peak position and intensities related to the deuterium desorption is visible in the 3D plot of the NPD diagrams (Figure 6). The 2D projection of the XRD and NPD patterns heated above 410 K (Figure 10) shows also clearly the shift of the peak positions toward a larger angle indicating a reduction in cell parameter due to D desorption. Note that a new peak at 2θ ≈ 55 ° appears above 420 K and its intensity increase can be related to the D content decrease. Both XRD and NPD diagrams clearly show a two-phase range around 500 K. This range corresponds to the second peak of the DSC and TDS curves. Cell volume and weight percentage variations were obtained by refining the XRD and NPD patterns, and are shown in Figures 11 and 12, respectively. Both the cell volume variation and the thermal desorption curves indicate that deuterium desorption begins at 400 K and ends at 550 K.

The XRD data analysis reveals that there are at least five different cubic phases (referred to as C$n$, where $n$ = 1 to 5), which are separated by temperature ranges where two deuterium containing phases are coexisting. The largest volume difference occurs between C4 and C5. The C5 phase exhibits the smallest unit cell and can be considered as an $\alpha$-phase (deuterium solid solution in the $TbFe_2$ intermetallic).

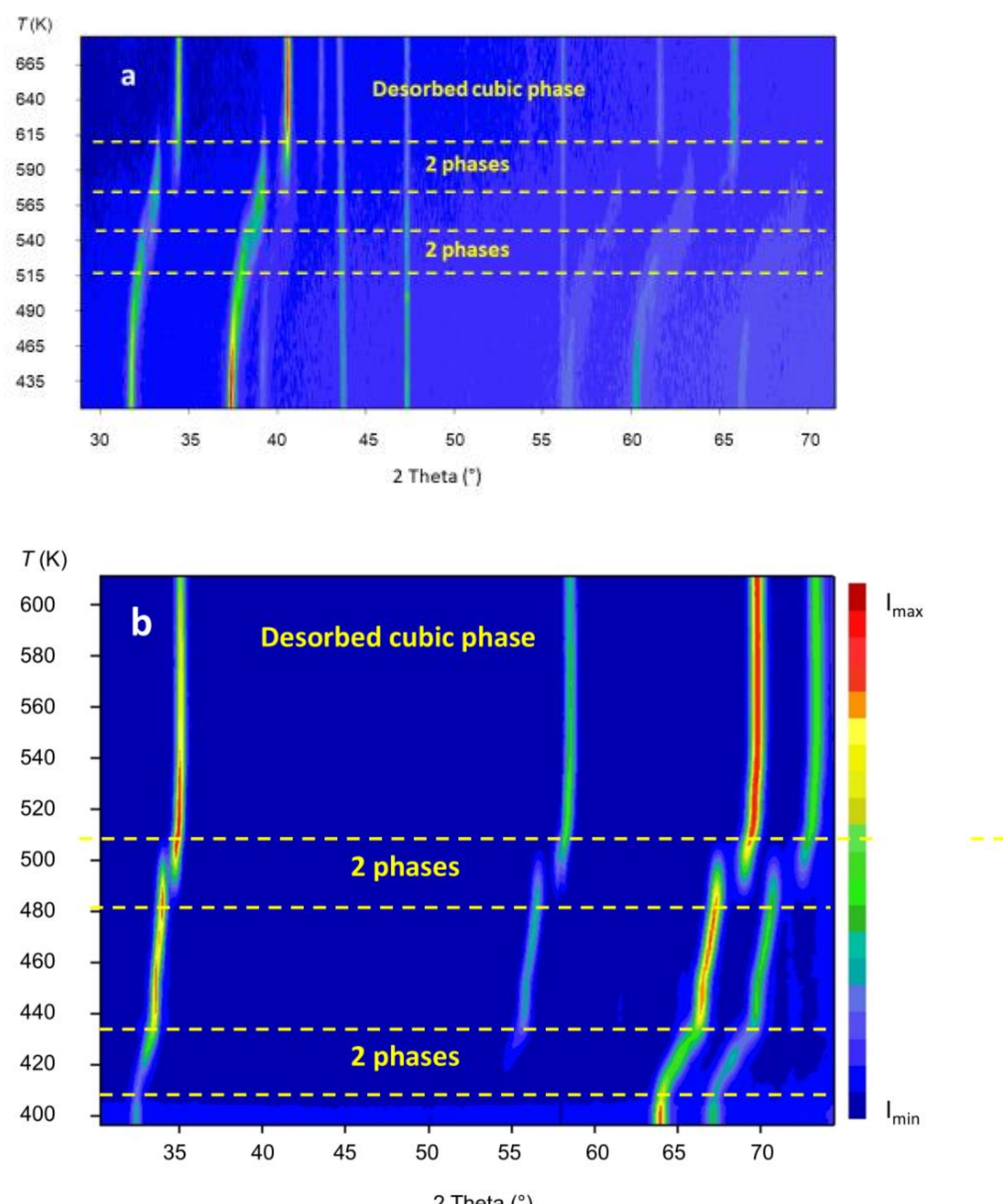

**Figure 10**: 2D plot of a) the XRD pattern of $TbFe_2D_{4.2}$ upon heating between 413 and 673 K and b) NPD patterns of $TbFe_2D_{4.2}$ (D1B) upon heating between 400 K and 610 K.

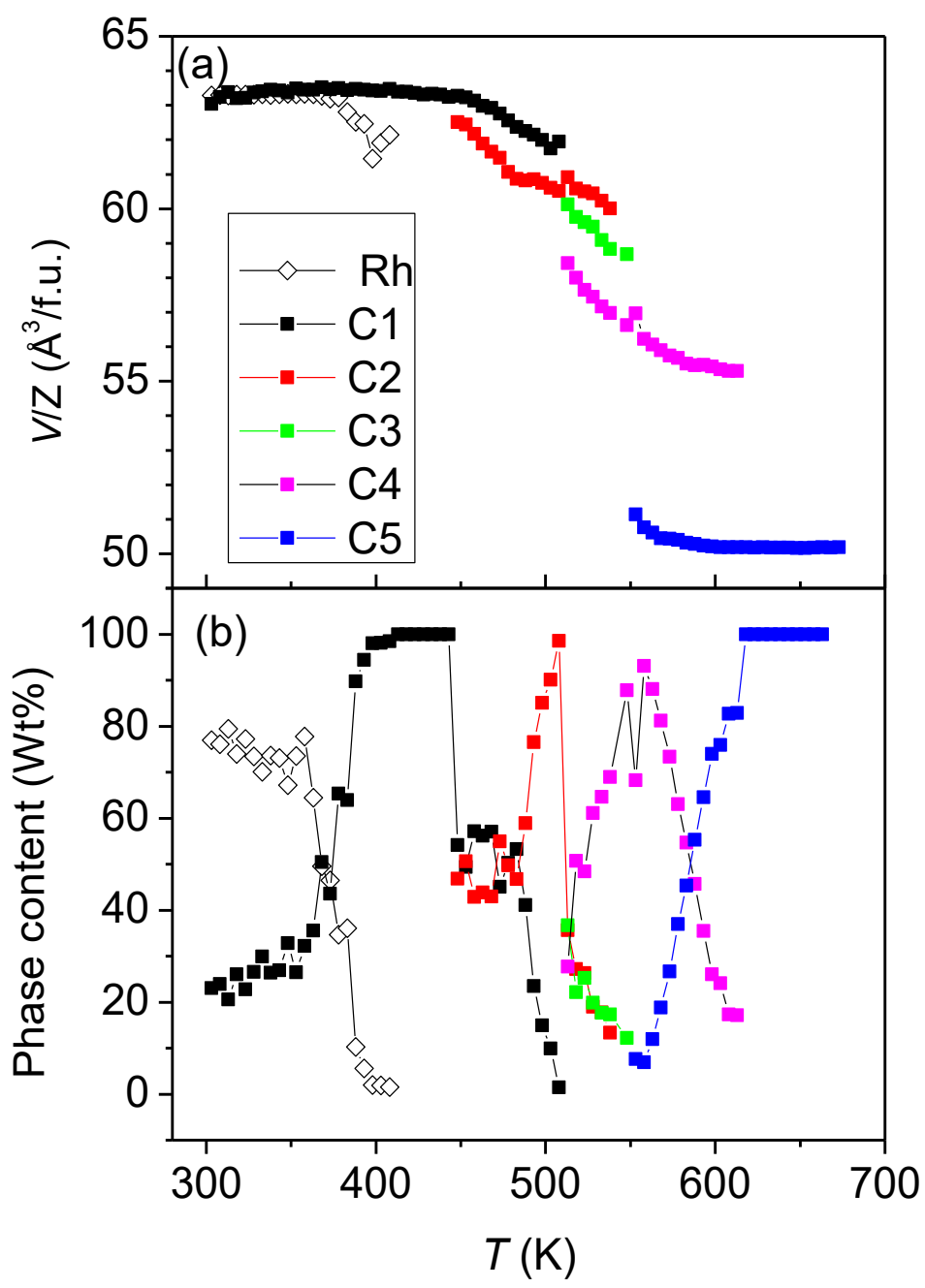

**Figure 11**: Thermal evolution of a) the volume by formula unit and b) the weight percent of the different phases observed during heating as derived from XRD patterns. Rh corresponds to the rhombohedral phase and C1 to C5 corresponds to the different cubic phases (*x* the D content being lower from C1 to C5) observed upon heating above 300 K and related to deuterium desorption from $TbFe_2D_{4.2}$.

In the monoclinic structure the D atoms are inserted in 13 [$Tb_2Fe_2$] sites and 5 [$TbFe_3$] sites. Upon heating the transformation toward a cubic structure in *Fd*-3*m* space group reduces the number of sites to 1 [$TbFe_2$] site and 1 [$TbFe_3$] site with a statistical D occupation. The refinement of the NPD pattern at 404 K with one single cubic phase shows that 3.64(8) D/f.u. atoms are in the [$Tb_2Fe_2$] site and 0.55(8) D/f.u. in [$TbFe_3$], corresponding to a total of 4.2 D/f.u. in agreement with the initial D content. Previous works on isostructural $YFe_2D_{4.2}$ has shown that even if the long range D order is lost, the local order is maintained at the scale of one unit cell ($d$ =8 Å) [7].

A close examination of the NPD patterns of $TbFe_2D_x$ measured upon heating reveals the presence of weak additional peaks between 435 and 462 K indicated by arrows in Figure 6. Similar superstructure peaks were observed upon D desorption from $YFe_2D_{4.2}$ in the NPD patterns between 435 and 460 K and attributed to $YFe_2D_{1.9}$. These superstructure peaks are attributed to the same lowering of crystal symmetry in the SR-XRD pattern of $TbFe_2D_{2.30}$. The pattern matching refinement of the $TbFe_2D_x$ NPD pattern at 446 K with $a_{tetra}$ = 12.157(1) Å and $c_{tetra}$ = 23.077(3) is presented in supplementary materials (Figure S6). On the other hand, no superstructure peaks corresponding to a cubic phase with a doubling of the *a* cell parameter as for $YFe_2D_{1.75}$ could be observed [5].

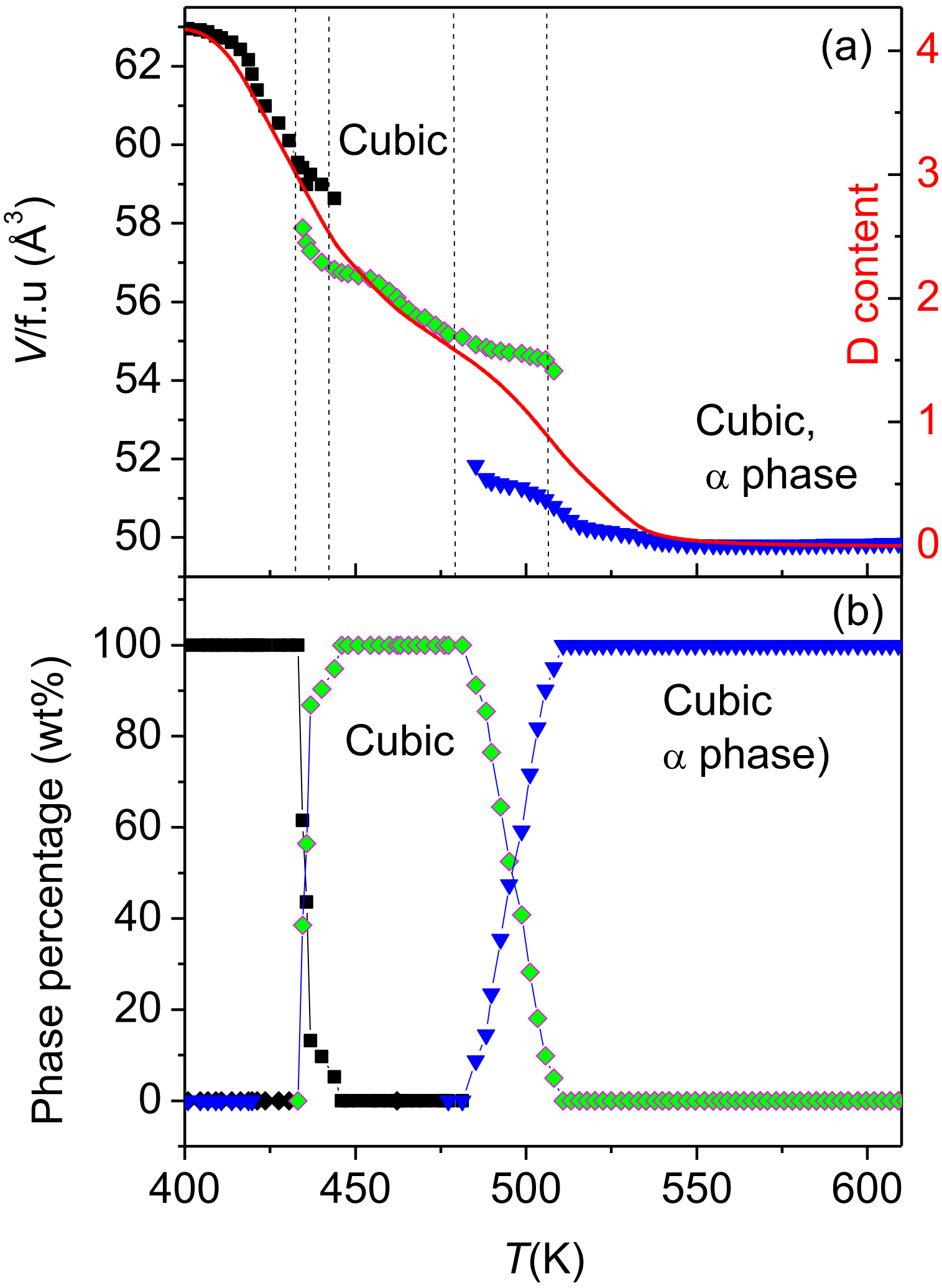

**Figure 12**: Thermal evolution of (a) the volume by formula unit and (b) weight percentage measured for the deuteride $TbFe_2D_{4.2}$ as derived from *in situ* NPD (D1B patterns) analysis of the thermal desorption.

This thermal and structural study has revealed that a reversible and first-order transition from monoclinic to cubic structure is observed above room temperature for $TbFe_2D_{4.2}$ similarly to what has been reported for $YFe_2D_{4.2}$ [7, 27]. These results show also that, as for $YFe_2$, deuterium insertion in $TbFe_2$ induces the formation of several phases with different structures derived from the initial C15 phase. Most of the deuterides have a cubic structure with a cell volume increase proportional to the D content ($\Delta V/V$). For specific D contents ($3.69 < x < 3.76$ and for $x = 4.2(1)$), a monoclinic distortion is observed at room temperature. In the *in-situ* measurements above 400 K, most of the deuterides which are above the O-D transition crystallize in a cubic structure. However, the presence of superstructure peaks, as for $TbFe_2D_{2.05}$, indicates the presence of different long-range ordered structures. In the case of $YFe_2D_x$ three different superstructure for x =1.3, 1.75 and 1.9 have been observed [5]. This probably also means different D-metal interactions depending on D content in $TbFe_2$.

It is surprising also that for a similar cubic $MgCu_2$ type structure with partial occupation of D atom in [$Tb_2Fe_2$] and [$TbFe_3$] sites, different cubic phases with different cell parameters are formed versus D content instead of a solid solution of D in one cubic phase with a continuous cell parameter decrease. But we have observed by XRD and NPD upon D desorption during heating between 400 and 550 K that all these deuterides exists for a narrow D content and are separated by two-phase ranges.

These results can explain the literature data, where either cubic or rhombohedral crystalline $TbFe_2$ hydrides were reported, as the structure strongly varies with the H (D) content. In their earlier study, Berthier et al. [15] found cubic hydrides for low H content and rhombohedral hydrides for $2.65 \leq x \leq 3.47$ on the basis of low resolution labscale X-ray diffractometer. Even considering that this rhombohedral is equivalent to the first monoclinic phase ($3.69 < x < 3.76$), the H concentration range is lower than in our study. This can originate from a problem of determining precisely the H content inside their samples. Mushnikov et al. [17], found also a rhombohedral distortion for $TbFe_2H_{3.56}$, whereas the other hydrides were cubic, in better agreement with the present results. As they both used laboratory scale XRD, the monoclinic distortion could not be clearly observed in the diffraction patterns, data were simply refined in rhombohedral structure.

The reversible order-disorder transition observed upon heating for $TbFe_2D_{4.2}$ between 355 and 380 K by NPD occurs at a slightly higher temperature than for $YFe_2D_{4.2}$ (between 345 and 360 K) [36]. Such order-disorder transition was also observed in $Y_{0.9}Gd_{0.1}Fe_2(H,D)_{4.2}$ compounds and the transition temperatures were found sensitive to the (H,D) isotope effect [32].

The multipeak thermal desorption behavior observed for both $YFe_2D_{4.2}$ [35] and $TbFe_2D_{4.2}$ can be attributed to the different transitions between two deuterides with different D contents, in agreement with the specific phase diagram of these Laves phase compounds, which contains several phases with different stability and therefore desorption temperatures.

The ability to form hydrides with structural distortions derived from their parent compound below an order-disorder temperature is a common property of many $RM_2$ Laves phase compounds ($R$ = Zr, Hf, rare earth; $M$ = V, Cr, Mn, Fe, Co) [4].

## 4. Conclusions

X-ray diffraction measurements including high resolution synchrotron data has been combined with neutron diffraction experiments, and calorimetry experiments to investigate the crystal structures of $TbFe_2D_x$ compounds for $x \leq 4.5$, and the evolution of $TbFe_2D_{4.2}$ versus temperature. $TbFe_2D_{4.2}$ crystallizes in a monoclinic structure (*Pc* space group), which is derived from the $MgCu_2$ cubic structure and isostructural to $YFe_2D_{4.2}$. The analysis of XRD patterns measured with synchrotron radiation for selected samples at room temperature confirms that D order yields the formation of at least 12 different deuterides separated by two-phase ranges as was previously observed for $YFe_2D_x$ compounds. A lowering of crystal symmetry in a tetragonal structure is observed for $TbFe_2D_{2.05}$ and monoclinic structures for $TbFe_2D_x$ deuterides with $x$= 3.7±0.1 and 4.1(1) D/f.u.. The monoclinic distortion increases with the D content. Upon heating, $TbFe_2D_{4.2}$ undergoes a reversible first-order monoclinic-cubic transition related to a transition from a long-range ordered structure to a disordered one and explained by D ordering. Then above 410 K, a deuterium desorption occurs through a multipeak behavior observed by DSC and thermal desorption curve. The XRD and NPD pattern analysis shows the existence of several two-phase ranges attributed to the co-existence of phases with different cell parameters and crystal structure symmetries. Superstructure peaks observed between 435 and 460 K can be attributed to a tetragonal phase with $x$=2.0. The atomic positions of the Tb and Fe atoms are reported from XRD.

This work allows us to propose several new crystal structures (tetragonal and monoclinic) and clarify the structural phase diagram of $TbFe_2D_x$ compounds. Further studies related to the magnetic properties of $TbFe_2D_{4.2}$ will be performed.

**Acknowledgments**

We thank E. Leroy (ICMPE laboratory) for the EPMA measurements on $TbFe_2$ alloy. We are thankful to T. Leblond and F. Cuevas for their help to measure the NPD patterns at ILL on D1B diffractometer. Some NPD measurements presented here are part of the PhD thesis of T. Leblond. We are grateful to the Laboratoire Léon Brillouin (3T2 instrument) and the Institut Laue Langevin (D1B instrument, proposal CRG-939) for the time allocated to perform neutron diffraction experiments.

We acknowledge the European Synchrotron Radiation Facility (ESRF) for provision of synchrotron radiation facilities and Momentum Transfer for facilitating the measurements. Jakub Drnec is thanked for assistance and support in using beamline ID31. The measurement setup was developed with funding from the European Union's Horizon 2020 research and innovation program under the STREAMLINE project (grant agreement ID 870313). Measurements performed as part of the MatScatNet project were supported by OSCARS through the European Commission's Horizon Europe Research and Innovation programme under grant agreement No. 101129751."

**References**

[1] A.E. Clark, Magnetostrictive rare earth-$Fe_2$ compounds, in: K.H.J. Buschow, E.P. Wohlfarth (Eds.) Handbook of Ferromagnetic Materials, Elsevier, North-Holland, Amsterdam 1980, pp. 531-589.

[2] N.C. Koon, C.M. Williams, B.N. Das, Giant magnetostriction materials, J. Magn. Magn. Mat., 100 (1991) 173-185.https://doi.org/10.1016/0304-8853(91)90819-V

[3] G. Wiesinger, G. Hilscher, Magnetism of hydrides, in: K.H.J. Buschow (Ed.) Handbook of Magnetic Materials, Elsevier North-Holland, Amsterdam, 2008, pp. 293-456

[4] H. Kohlmann, Hydrogen order in hydrides of Laves phases, Z. Krist-Cryst Mater., 235 (2020) 319-332.https://doi.org/10.1515/zkri-2020-0043

[5] V. Paul-Boncour, L. Guénée, M. Latroche, A. Percheron-Guégan, B. Ouladdiaf, F. Bourée-Vigneron, Elaboration, structures and phase transitions for $YFe_2D_x$ compounds (x=1.3, 1.75, 1.9, 2.6) studied by neutron diffraction., J. Solid State Chem., 142 (1999) 120-129.https://doi.org/10.1006/jssc.1998.7995

[6] V. Paul-Boncour, M. Latroche, A. Percheron-Guégan, O. Isnard, Study of phase transformations in $YFe_2D_{1.75}$ deuterides by in situ neutron diffraction, Physica B, 276-278 (2000) 278-279.https://doi.org/10.1016/S0921-4526(99)01461-1

[7] J. Ropka, R. Cerný, V. Paul-Boncour, Local deuterium order in apparently disordered Laves phase deuteride $YFe_2D_{4.2}$, J. Solid State Chem., 184 (2011) 2516-2524.https://doi.org/10.1016/j.jssc.2011.07.028

[8] V. Paul-Boncour, S.M. Filipek, A. Percheron-Guégan, I. Marchuk, J. Pielaszek, Structural and magnetic properties of $RFe_2H_5$ hydrides (R=Y,Er), J. Alloys Compds, 317-318 (2001) 83-87.https://doi.org/10.1016/S0925-8388(00)01400-6

[9] V. Paul-Boncour, S.M. Filipek, I. Marchuk, G. André, F. Bourée, G. Wiesinger, A. Percheron-Guégan, Structural and magnetic properties of $ErFe_2D_5$ studied by neutron diffraction and Mössbauer spectroscopy, J. Phys.: Condens. Matter, 15 (2003) 4349-4359.http://doi.org/10.1088/0953-8984/15/25/306

[10] M. Caussé, G. Geneste, L. Toraille, B. Guigue, J.B. Charraud, V. Paul-Boncour, P. Loubeyre, Synthesis of Laves phase hydrides $YFe_2H_6$ and $YFe_2H_7$ at high pressure: Reaching a limit of interstitial hydrogen uptake, J. Alloys Compds, 1010 (2025) 9 https://doi.org/10.1016/j.jallcom.2024.177392

[11] K. Aoki, K. Mori, H. Onodera, T. Matsumoto, Hydrogen-induced amorphization of C15 Laves $TbFe_2$ compound, J. Alloys Compds, 253-254 (1997) 106-109.https://doi.org/10.1016/S0925-8388(96)02909-X

[12] K. Aoki, M. Dilixiati, K. Ishikawa, Hydrogen-induced transformations in C15 Laves phases $CeFe_2$ and $TbFe_2$ studied by pressure calorimetry up to 5 MPa, J. Alloys Compds, 356-357 (2003) 664-668.https://doi.org/10.1016/S0925-8388(03)00148-8

[13] N.K. Zajkov, N.V. Mushnikov, V.S. Gaviko, A.Y. Yermakov, Effect of high-temperature hydrogen treatment on magnetic properties and structure of $TbFe_2$-based compounds, Int. J. Hydr. Energ., 22 (1997) 249-253.https://doi.org/10.1016/s0360-3199(96)00168-1

[14] K. Itoh, K. Kanda, K. Aoki, T. Fukunaga, X-Ray and neutron diffraction studies of atomic scale structures of crystalline and amorphous $TbFe_2D_x$, J. Alloys Compds, 348 (2003) 167-172.https://doi.org/10.1016/S0925-8388(02)00842-3

[15] Y. Berthier, T. de Saxce, D. Fruchart, P. Vulliet, Magnetic interactions and structural properties of ternary hydrides $TbFe_2H_x$, Physica B, 130B (1985) 520-523 https://doi.org/10.1016/0378-4363(85)90293-1

[16] S.K. Kulshreshtha, O.D. Jayakumar, R. Sasikala, Hydrogen-induced spin reorientation in $TbFe_2H_x$ system, J. Magn. Magn. Mat., 117 (1992) 33-37.https://doi.org/10.1016/0304-8853(92)90288-y

[17] N.V. Mushnikov, N.K. Zaikov, V.S. Gaviko, A.V. Korolev, A.E. Ermakov, Hydrogen-induced magnetic anisotropy and magnetostriction in rare-earth intermetallics with $MgCu_2$-type structure, Russian Metallurgy, (1995) 86-91.https://doi.org/10.1134/S0031918X13120077

[18] T. Leblond, V. Paul-Boncour, M. Guillot, O. Isnard, Metamagnetic transitions in $RFe_2(H,D)_{4.2}$ compounds (R=Y, Tb), J. Appl. Phys., 101 (2007) 09G514.https://doi.org/10.1063/1.2710456
[19] V. Paul-Boncour, M. Guillot, O. Isnard, A. Hoser, High field induced magnetic transitions in the $Y_{0.7}Er_{0.3}Fe_2D_{4.2}$ deuteride, Phys. Rev. B, 96 (2017) 104440.https://doi.org/10.1103/PhysRevB.96.104440
[20] V. Paul-Boncour, O. Isnard, M. Guillot, A. Hoser, Metamagnetic transitions in $Y_{0.5}Er_{0.5}Fe_2D_{4.2}$ deuteride studied by high magnetic field and neutron diffraction experiments, J. Magn. Magn. Mat., 477 (2019) 356-365.https://doi.org/10.1016/j.jmmm.2019.01.056
[21] V. Paul-Boncour, O. Isnard, V. Shtender, Y. Skourski, M. Guillot, Origin of the metamagnetic transitions in $Y_{1-x}Er_xFe_2(H,D)_{4.2}$ compounds, J. Magn. Magn. Mat., 512 (2020) 167018.https://doi.org/10.1016/j.jmmm.2020.167018
[22] V. Paul-Boncour, A. Herrero, V. Shtender, K. Provost, E. Elkaim, Magnetic transitions with magnetocaloric effects near room temperature related to structural transitions in $Y_{0.9}Pr_{0.1}Fe_2D_{3.5}$ deuteride, J. Appl. Phys., 130 (2021) 113904.https://doi.org/10.1063/5.0061200
[23] V. Paul-Boncour, K. Provost, E. Alleno, A. N'Diaye, F. Couturas, E. Elkaim, Phase diagram and order-disorder transitions in $Y_{0.9}Gd_{0.1}Fe_2H_x$ hydrides ($x \geq 2.9$), J. Alloys Compds, 896 (2022) 163016.https://doi.org/10.1016/j.jallcom.2021.163016
[24] J. Rodriguez-Carvajal, Recent Advances in Magnetic Structure Determination by Neutron Powder Diffraction, Physica B, 192 (1993) 55-69.https://doi-org.inc.bib.cnrs.fr/10.1016/0921-4526(93)90108-I
[25] A.E. Dwight, C.W. Kimball, $TbFe_2$, a rhombohedral Laves phase, Acta Crystallographica Section B, 30 (1974) 2791-2793.doi:10.1107/S0567740874008156
[26] M.P. Dariel, J.T. Holthuis, M.R. Pickus, The terbium-iron phase diagram, J. Less-Common Met., 45 (1976) 91-101.https://doi.org/10.1016/0022-5088(76)90200-9
[27] J. Ropka, R. Cerny, V. Paul-Boncour, T. Proffen, Deuterium ordering in Laves phases deuterides $YFe_2D_{4.2}$, J. Solid State Chem., 182 (2009) 1907-1912.https://doi.org/10.1016/j.jssc.2009.04.033
[28] Z. Arnold, O. Isnard, V. Paul-Boncour, Influence of high pressure on the remarkable itinerant electron behavior in $Y_{0.7}Er_{0.3}Fe_2D_{4.2}$ compound, J. Appl. Phys., 133 (2023) 173901.https://doi.org/10.1063/5.0141855

[29] V. Paul-Boncour, V. Shtender, K. Provost, M. Phejar, F. Cuevas, Y. Skourski, O. Isnard, Origin of the metamagnetic transitions in $Y_{0.9}Tb_{0.1}Fe_2D_{4.3}$, J. Solid State Chem., 338 (2024) 9.https://doi.org/10.1016/j.jssc.2024.124898

[30] D.G. Westlake, Site occupancies and stoichiometries in hydrides of intermetallic compounds : geometric considerations, J. Less-Common Met., 90 (1983) 251-273

[31] A.C. Switendick, Band structure calculations for metal hydrogen systems, Z. Phys. Chem. Neue Fol., 117 (1979) 89-112.https://doi.org/10.1524/zpch.1979.117.117.089

[32] V. Paul-Boncour, S. Voyshnis, K. Provost, J.C. Crivello, Isotope effect on structural transitions in $Y_{0.9}Gd_{0.1}Fe_2(H_zD_{1-z})_{4.2}$ compounds, Chem. Met. Alloys, 6 (2013) 130-143.https://doi.org/10.30970/cma6.0249

[33] J. Ropka, Studies of local order in apparently disordered hydrides of Laves phases $YFe_2D_x$, $YMn_2D_x$, $ZrV_2D_x$ and of $La(Ni_{4:5}Sn_{0.5})D_{3.85}$, PhD thesis, Université de Genève, director R. Cerny

[34] Y.J. Tang, Transition-metal substitution effect on magnetic and magnetostrictive properties of $TbFe_2$ compounds, J. Magn. Magn. Mat., 167 (1997) 245-248.https://doi.org/10.1016/S0304-8853(96)00331-9

[35] T. Leblond, V. Paul-Boncour, F. Cuevas, O. Isnard, J.F. Fernandez, Study of the multipeak deuterium thermodesorption in $YFe_2D_x$ ($1.3 <= x <= 4.2$) by DSC, TD and in situ neutron diffraction, Int. J. Hydr. Energ., 34 (2009) 2278-2287.https://doi.org/10.1103/PhysRevB.84.094429

[36] O. Isnard, V. Paul-Boncour, Z. Arnold, C.V. Colin, T. Leblond, J. Kamarad, H. Sugiura, Pressure-induced changes in the structural and magnetic properties of $YFe_2D_{4.2}$, Phys. Rev. B, 84 (2011) 094429.https://doi.org/10.1103/PhysRevB.84.094429

## Graphical Abstract

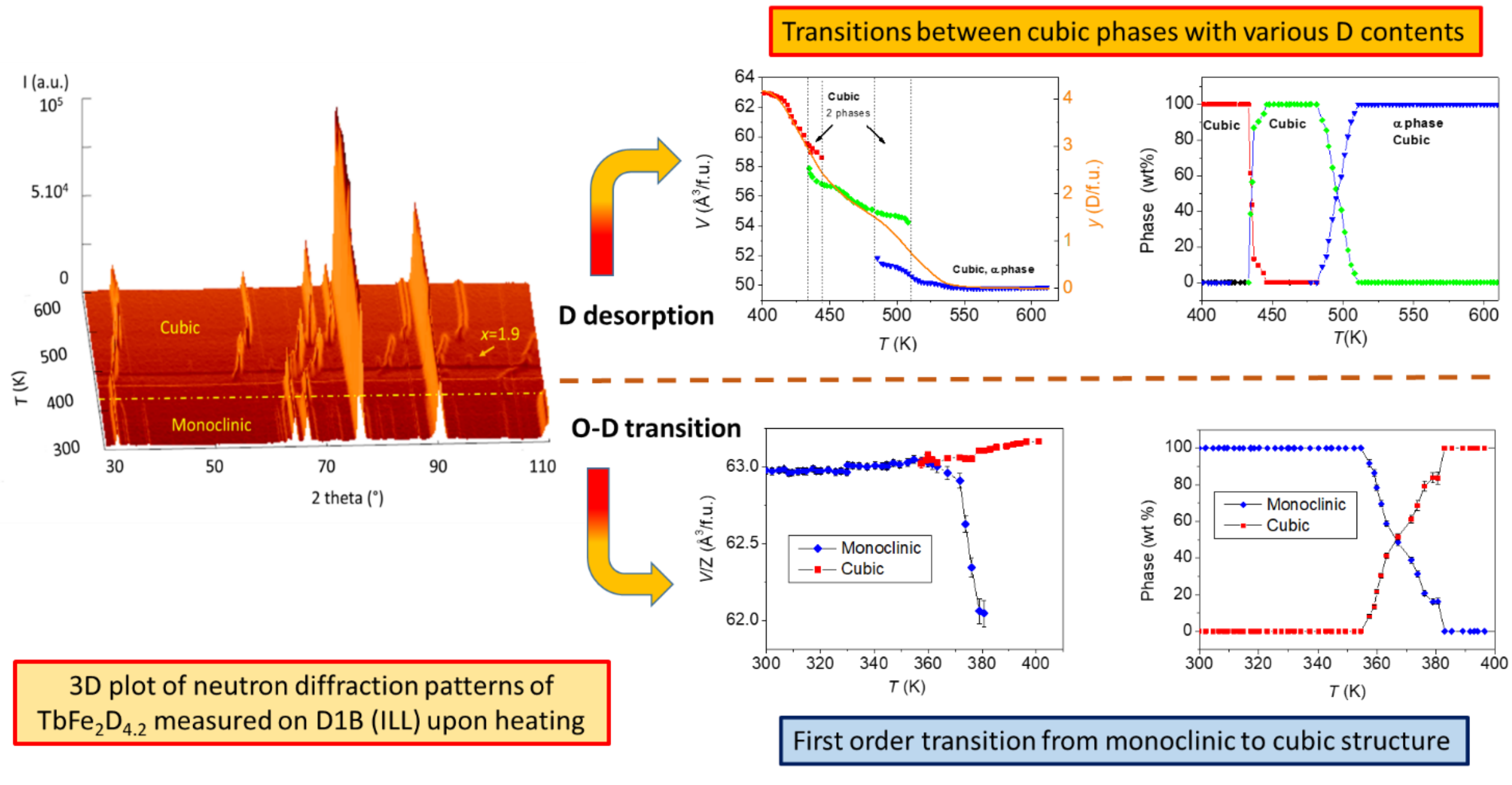

**Supplementary materials**

Structural transitions related to order-disorder and thermal desorption of D atoms in $TbFe_2D_{4.2}$

V. Paul-Boncour[1]*, O. Isnard[2]

[1] *Université Paris-Est Créteil, CNRS, ICMPE, UMR7182, F-94320 Thiais, France*

[2]*Université Grenoble Alpes, Institut Néel, CNRS, BP166X, 38042 Grenoble Cédex 9, France*

**Table S1**: Deuterium neighbors, total number of D neighbors considering the occupancy factors for distances between each metal atom ($M_i$) and each closest $D_j$ neighbors, individual $M_i$-$D_j$ and average $M_i$-D distances for $TbFe_2D_{4.2}$.

| Central atom ($M_i$) | D neighbors ($D_j$) | Total D neighbors | $d(M_i$-$D_j)$ (Å) | Average $d(M_i$-D) (Å) |
|---|---|---|---|---|
| Tb1 | D14<br>D9<br>D3<br>D5<br>D4<br>D7 | 7.3(1) | 2.03(4)<br>2.17(5)<br>2.29(3)<br>2.30(5)<br>2.33(4)<br>2.39(7) | 2.3(2) |
| Tb2 | D17<br>D1<br>D7<br>D10<br>D4<br>D8<br>D18<br>D2 | 6.7(1) | 1.83(5)<br>1.87(6)<br>2.12(6)<br>2.14(3)<br>2.21(5)<br>2.21(5)<br>2.44(4)<br>2.55(4) | 2.2(2) |
| Tb3 | D2<br>D11<br>D8<br>D13<br>D15<br>D6<br>D10<br>D1 | 7.4(1) | 1.98(3)<br>2.18(3)<br>2.22(4)<br>2.31(3)<br>2.34(3)<br>2.35(5)<br>2.39(4)<br>2.43(8) | 2.3(1) |
| Tb4 | D5<br>D13<br>D9<br>D6<br>D11<br>D12<br>D3<br>D14 | 7.6(1) | 2.07(4)<br>2.07(4)<br>2.18(5)<br>2.27(4)<br>2.28(4)<br>2.28(4)<br>2.31(5)<br>2.34(3) | 2.2(1) |
| Fe1 | D14 | 3.8(1) | 1.66(4) | 1.8(1) |

| | D16<br>D18<br>D13 | | 1.77(4)<br>1.77(5)<br>1.79(4) | |
|---|---|---|---|---|
| Fe2 | D7<br>D18<br>D5<br>D13<br>D9 | 4.5(1) | 1.60(6)<br>1.61(4)<br>1.68(4)<br>1.75(6)<br>1.90(4) | 1.6(1) |
| Fe3 | D3<br>D18<br>D11<br>D6<br>D1 | 4.4(1) | 1.61(4)<br>1.64(4)<br>1.74(6)<br>1.78(4)<br>1.84(6) | 1.7(1) |
| Fe4 | D12<br>D16<br>D2<br>D4<br>D7 | 4.4(1) | 1.54(6)<br>1.62(4)<br>1.74(3)<br>1.78(4)<br>1.80(6) | 1.7(1) |
| Fe5 | D8<br>D4<br>D6<br>D17<br>D12<br>D15 | 5.3(1) | 1.61(4)<br>1.64(4)<br>1.83(4)<br>2.15(5)<br>2.42(6)<br>2.52(4) | 2.0(4) |
| Fe6 | D15<br>D17<br>D10<br>D8<br>D2<br>D1 | 5.1(1) | 1.57(4)<br>1.71(5)<br>1.76(5)<br>1.79(4)<br>1.91(3)<br>2.00(6) | 1.8(2) |
| Fe7 | D14<br>D9<br>D3<br>D5<br>D16 | 5.0(1) | 1.63(4)<br>1.65(5)<br>1.71(4)<br>1.71(4)<br>1.93(4) | 1.7(1) |
| Fe8 | D15<br>D11<br>D12<br>D10<br>D17 | 4.3(1) | 1.60(4)<br>1.72(4)<br>1.72(5)<br>1.88(3)<br>2.30(6) | 1.8(2) |

The SR-XRD patterns were measured on ID31 synchrotron beamline at ESRF from Momentum transfer. The SR-XRD measurements were performed as part of the MatScatNet project were supported by OSCARS through the European Commission's Horizon Europe Research and Innovation programme under grant agreement No. 101129751.

The SR-XRD patterns of 4 $TbFe_2D_x$ ($x$= 2.30 to 3.55) were refined by the Rietveld method using the fullprof code. The cell parameters, weight percentage and agreement factors are given in the legend of Figures S1 to S4. The refinement of SR-XRD pattern of $TbFe_2D_{4.04}$ is in the main text. The results of the refinement of all samples are compared in table S1. (λ = 0.1652 Å)

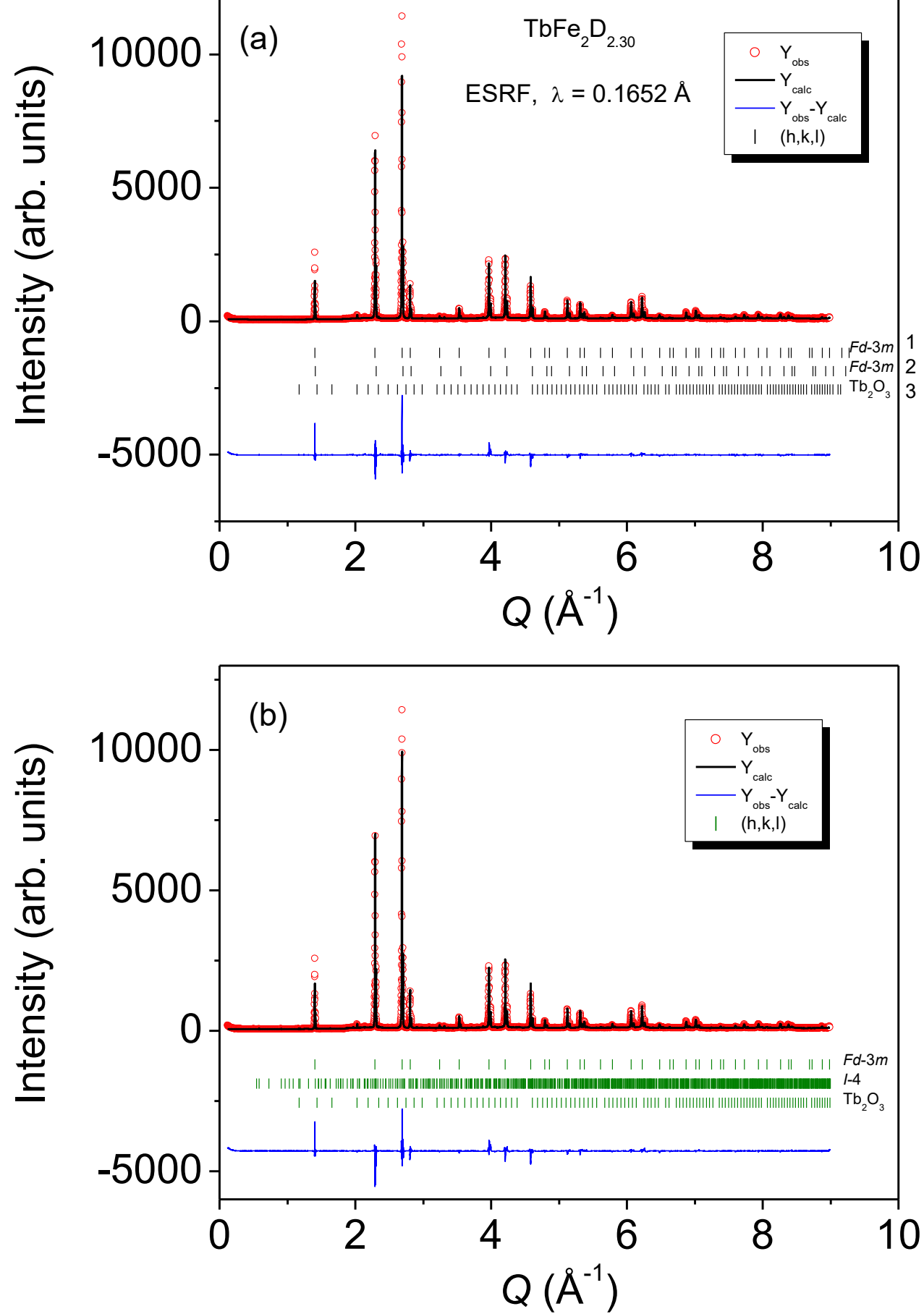

**Figure S1** : Refined SR-XRD pattern of $TbFe_2D_{2.30}$

(a/b) $R_p$: 7 % $R_{wp}$: 10.2 %, $R_p$: /5.7%, $R_{wp}$: 8.9%

Phase 1: $TbFe_2D_{2.39:}$ 85.1(4) wt% - cubic $a$= 7.7556(3) Å, $R_{Bragg}$: 6.5 %

/83.2 wt%, - cubic $a$= 7.75548 (3) Å $R_{Bragg}$: 6 %

Phase 2: $TbFe_2D_{2.05}$ 16.2(1) wt% - cubic $a$= 7.70901(4) Å, $R_{Bragg}$: 22 % /

14.3(1) wt% - tetragonal $a$= 12.1864(2), c= 23.1368(7) Å $R_{Bragg}$: 45 %

Phase 3: $Tb_2O_3$ 0.6(1) wt%- cubic $a$= 10.7326(3) Å, $R_{Bragg}$: 90 %

0.6(1) wt%- cubic $a$= 10.7325(3) Å, /, $R_{Bragg}$: 76 %

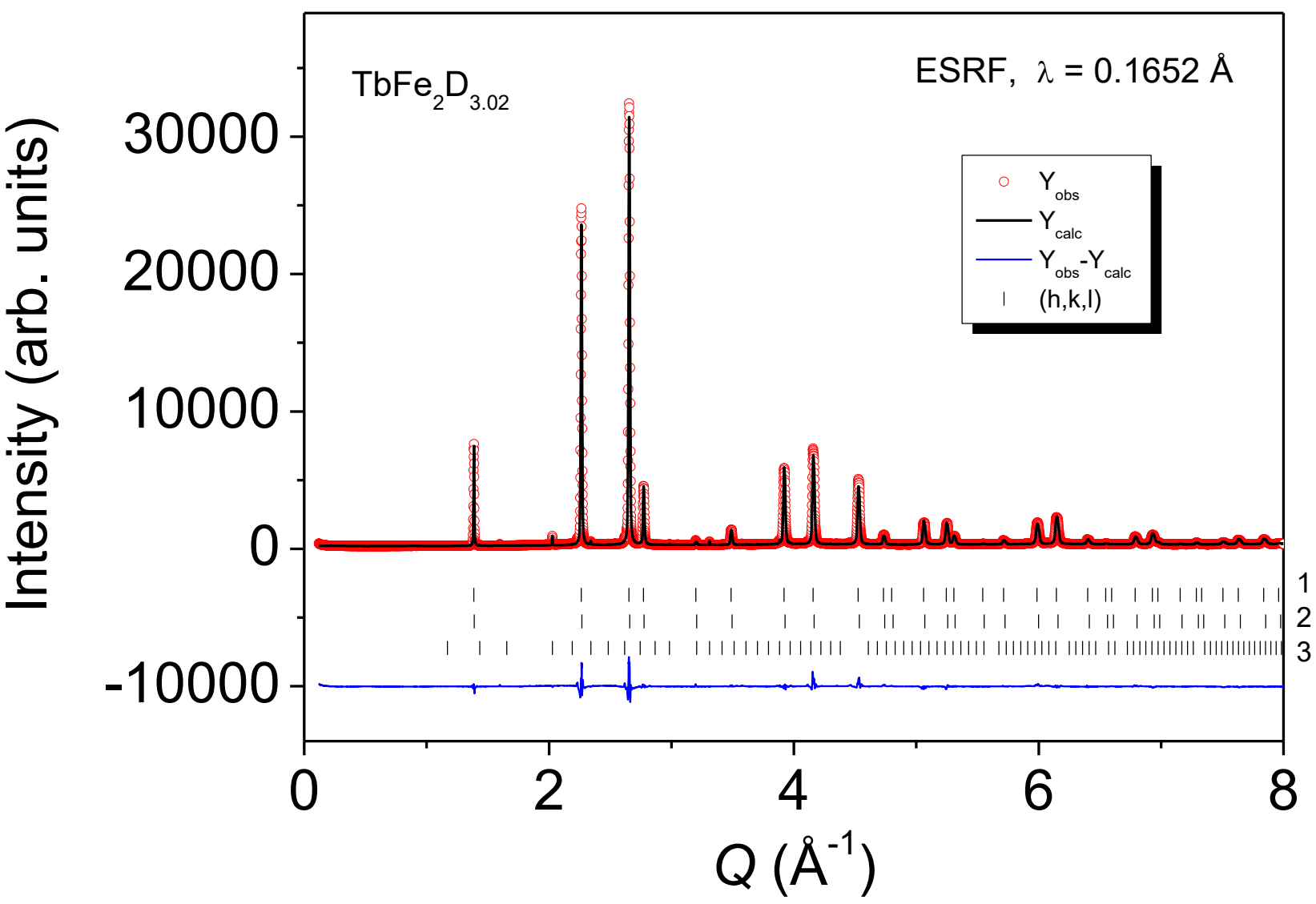

**Figure S2** : Refined SR-XRD pattern of $TbFe_2D_{3.02}$, $R_p$: 5.25 % $R_{wp}$: 7.61 %

Phase 1: $TbFe_2D_x$, 55.2(7) wt% - cubic $a$= 7.85423(7) Å- $R_{Bragg}$: 11.3 %

Phase 2: $TbFe_2D_x$, 44.0(1) wt% - cubic $a$= 7.8372(1) Å, $R_{Bragg}$: 13.1 %

Phase 3: $Tb_2O_3$, 0.8(1) wt% - cubic $a$= 10.7347(2) Å, $R_{Bragg}$: 50%

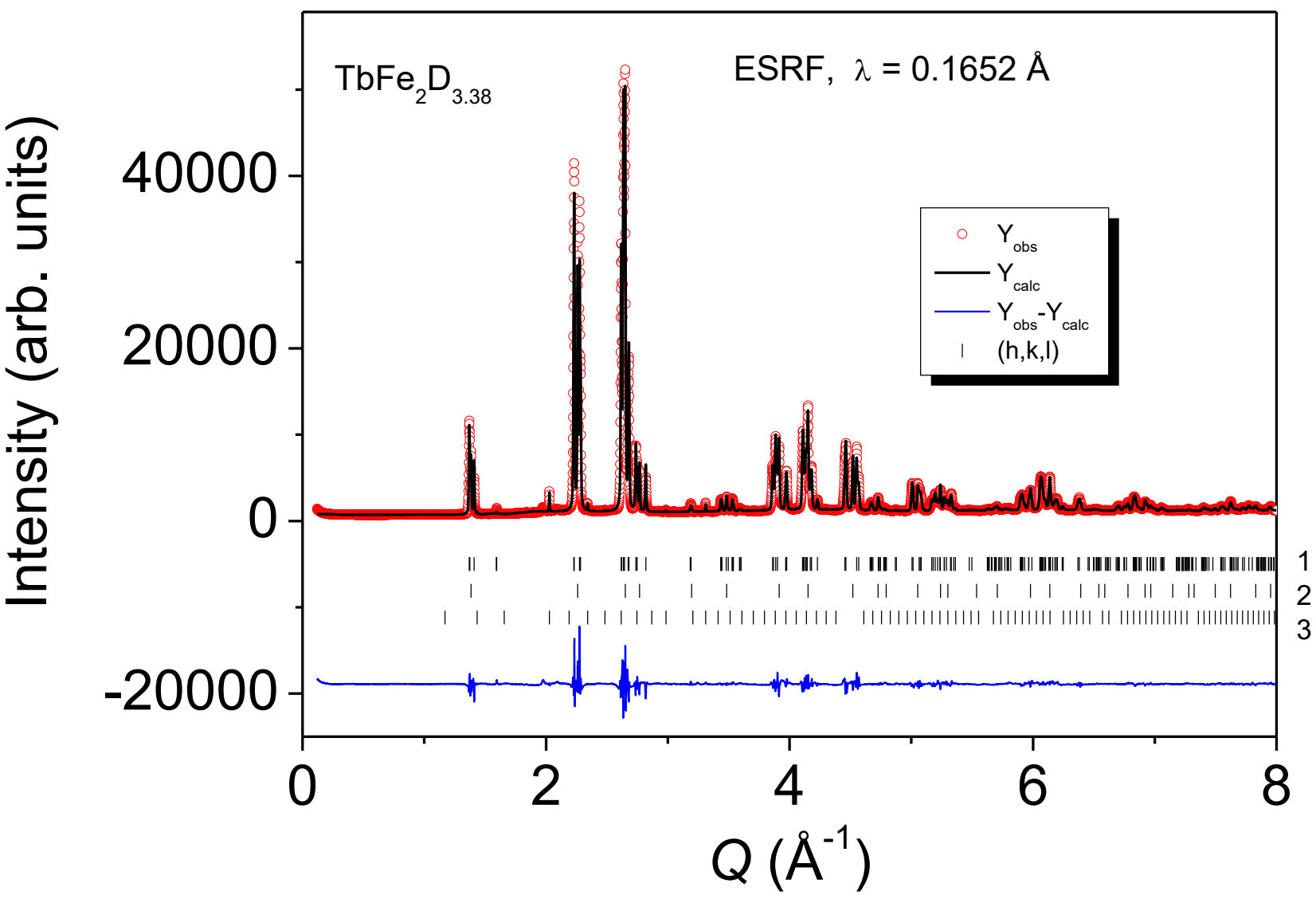

**Figure S3**: Refined SR-XRD pattern of $TbFe_2D_{3.38}$, $R_p$: 5.0 %, $R_{wp}$: 0 %

Phase 1: $TbFe_2D_{3.45}$, 76.8 (4) wt% - Monoclinic $a$= 9.50190(5) Å, $b$=5.64131(4) Å, $c$=5.50406(3) Å, $\beta$=123.8308(5) (°) - $R_{Bragg}$: 4.7 %

Phase 2: $TbFe_2D_{3.25}$, 22.5 (2) wt% - cubic $a$= 7.86292(5) Å, $R_{Bragg}$: 3.4 %

Phase 3: $Tb_2O_3$, 0.7(1) wt% - cubic $a$= 10.735(2) Å, $R_{Bragg}$: 25.3%

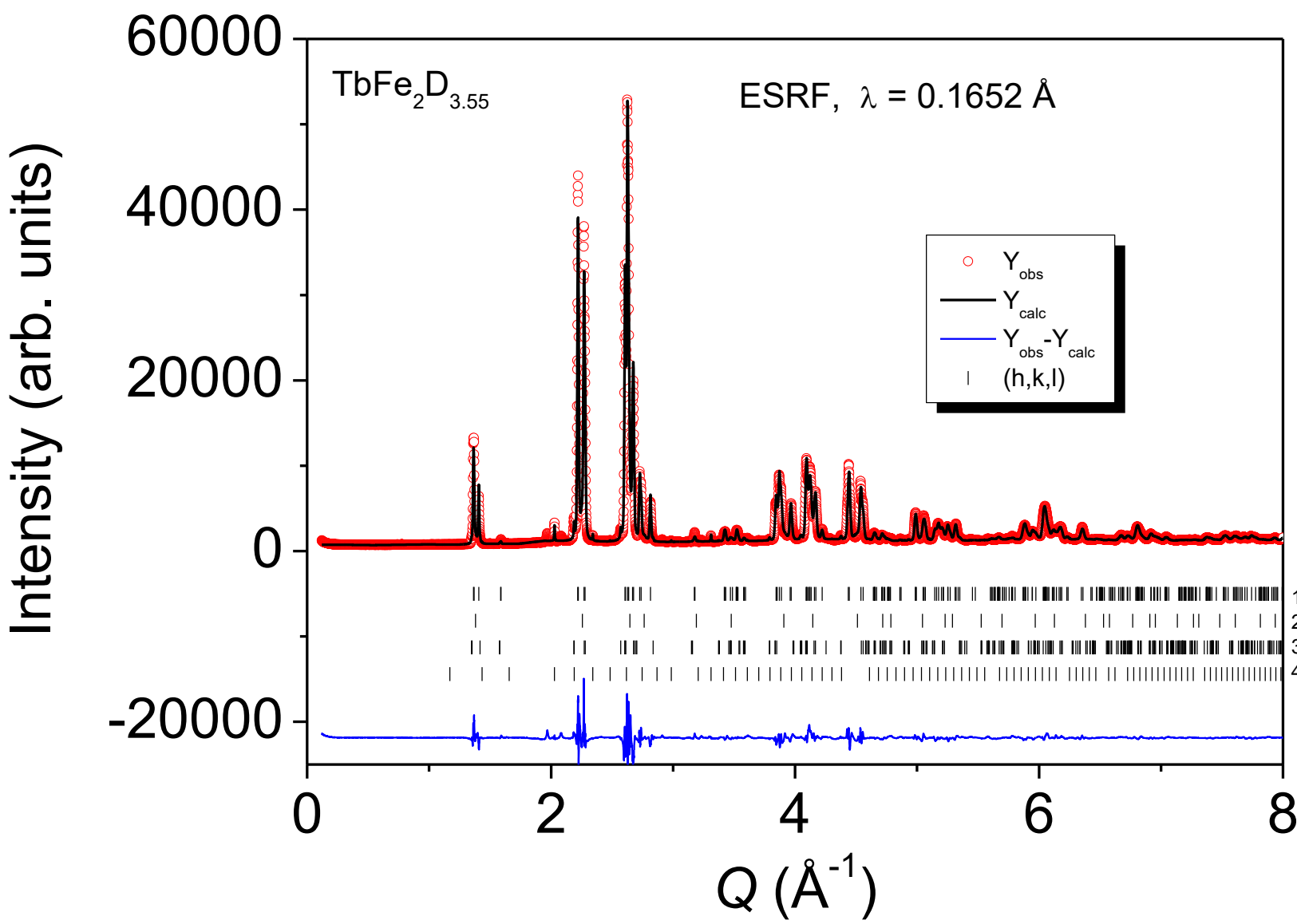

**Figure S4** : Refined SR-XRD pattern of $TbFe_2D_{3.55}$, $R_p$: 5.7 % $R_{wp}$: 7.8 %

Phase 1: $TbFe_2D_{3.69}$, 86.5(4) wt% - Monoclinic $a$= 9.52080 (8) Å, $b$= 5.66677 (7) Å, $c$=5 5.51620 (5) Å, $\beta$= 123.71956 (8) (°) - $R_{Bragg}$: 5.4 %

Phase 2: $TbFe_2D_{3.38}$, 10.3(5) wt% - cubic $a$= 7.8793 (4) Å, $R_{Bragg}$: 3.7%

Phase 3: $TbFe_2D_{4.3}$, 2.6 (5) wt% - Monoclinic $a$= 9.4396 (4) Å, $b$= 5.7503 (1) Å, $c$= 5.5232 (1) Å, $\beta$= 122.522 (2) (°) - $R_{Bragg}$: 13.1 %

Phase 4: $Tb_2O_3$, 0.6(1) wt% - cubic $a$= 10.735(2) Å, $R_{Bragg}$: 17.8 %

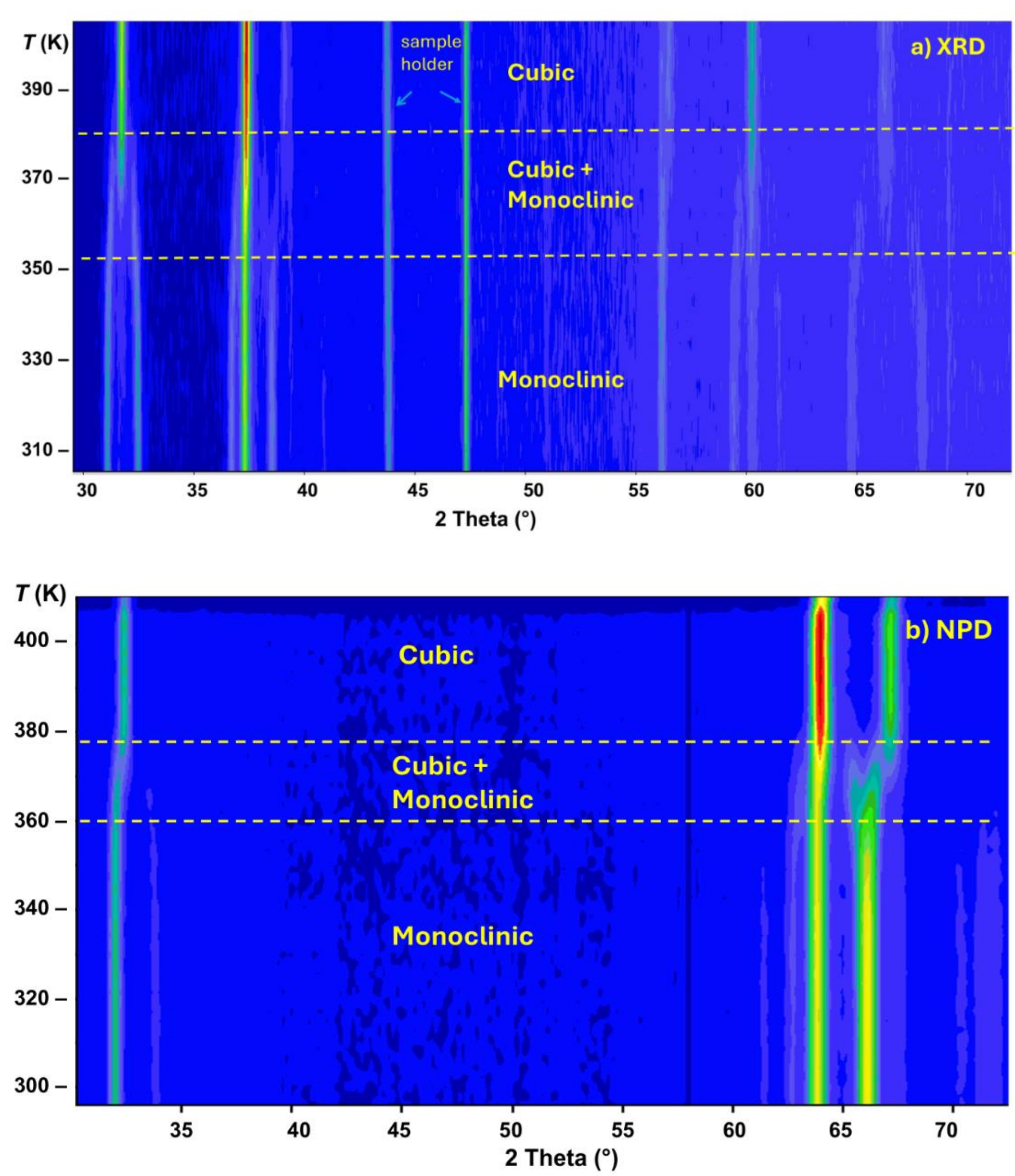

**Figure S5**: Evolution of a) the XRD ( $\lambda_{Cu\ K\alpha}$) and b) the NPD patterns (D1B, $\lambda$ = 2.52 Å) of $TbFe_2D_{4.2}$ upon heating between 300 and 410 K showing the transition from monoclinic to cubic structure. (blue= low intensity, red= higher intensity)

:

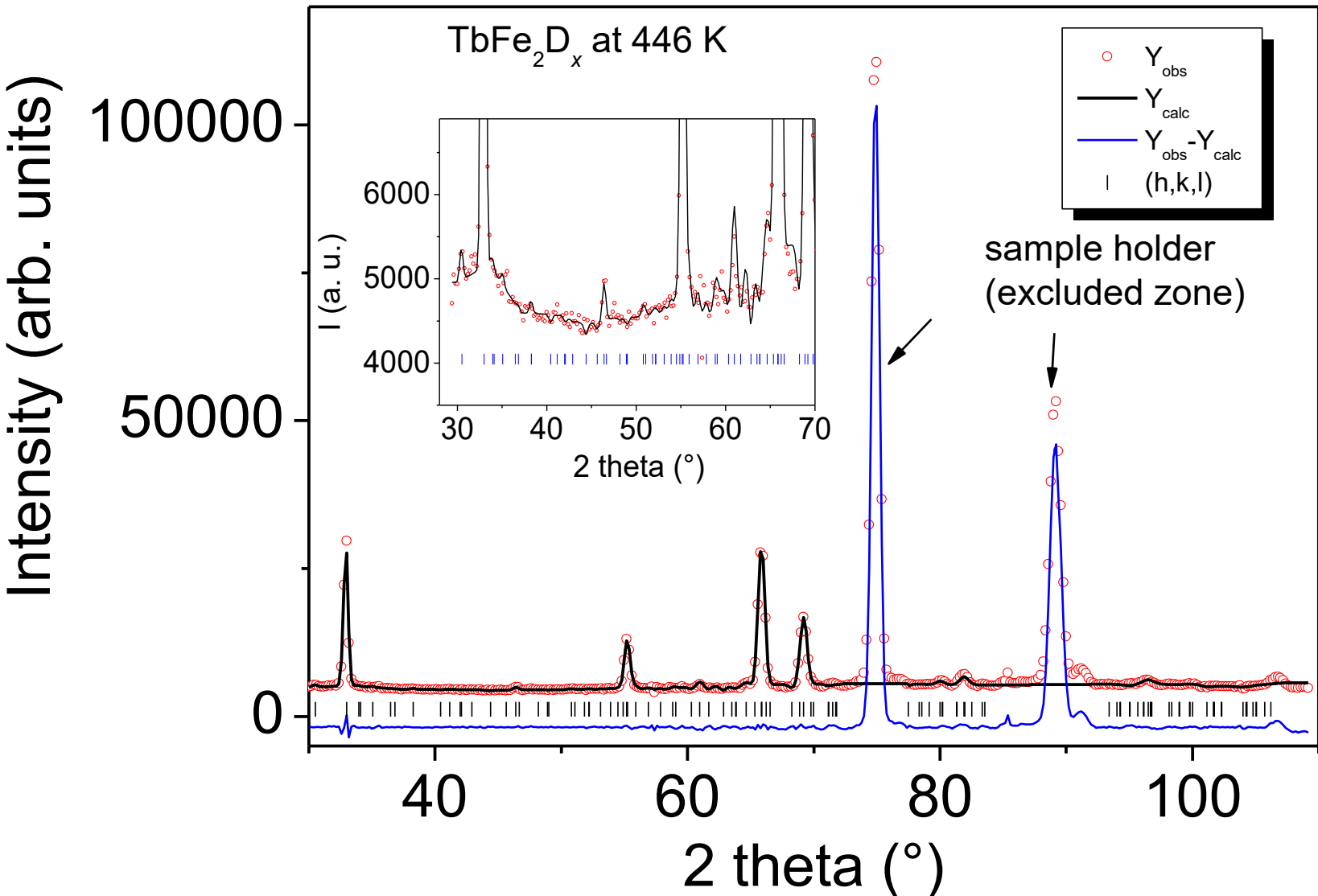

**Figure S6**: Experimental and refined NPD pattern using pattern matching method measured on D1B diffractometer at 446 K (λ = 2.52 Å) with a tetragonal cell corresponding to $TbFe_2D_{2.0}$. Inset: zoom showing the fit of superstructure peaks.

**Table S2**: D content, phase number, Space group, unit cell volume, cell volume/formula unit, weight percentage and calculated *x* content obtained from the Rietveld refinement of the SR-XRD patterns as displayed in figures S1 to S5. *x* (meas.) corresponds to the value of D inside the sample measured by a volumetric method, *x* (calc.) for each $TbFe_2D_x$ phase is calculated using the reduced cell volume (details in main text) and *x* (calc.) is the mean value for each sample using the weight percentage and the previous *x* for each phase.

| *x* (meas.) (D/f.u.) Sample | N° | phase | Space group | *V* (Å$^3$) | *V/Z* (Å$^3$/f.u.) | Wt % | *x* (Calc.) (D/f.u.) for each phase | *x* (Calc.) (D/f.u.) Mean value |
|---|---|---|---|---|---|---|---|---|
| 2.30(5) | 1 | $TbFe_2D_x$ | *I*-4 | 466.485(4) | 58.311(1) | 85.1(4) | 2.39 | 2.34 |
| | 2 | $TbFe_2D_x$ | *Fd*-3*m* | 458.171(5) | 57.271(1) | 14.3(1) | 2.07 | |
| | 3 | $Tb_2O_3$ | *Ia*-3 | 1236.27(7) | 77.267(1) | 0.6(1) | | |
| 3.02(5) | 1 | $TbFe_2D_x$ | *Fd*-3*m* | 484.52(1) | 60.565(1) | 55.2(7) | 3.07 | 3.05 |
| | 2 | $TbFe_2D_x$ | *Fd*-3*m* | 481.367(8) | 60.170(1) | 44.0(7) | 3.02 | |
| | 3 | $Tb_2O_3$ | *Ia*-3 | 1237.02(4) | 77.314(1) | 0.8(1) | | |
| 3.38(5) | 1 | $TbFe_2D_x$ | *C*2/*m* | 245.081(3) | 61.270(1) | 76.8(3) | 3.45 | 3.40 |
| | 2 | $TbFe_2D_x$ | *Fd*-3*m* | 486.129(2) | 60.766(1) | 22.5(2) | 3.25 | |
| | 3 | $Tb_2O_3$ | *Ia*-3 | 1236.604(4) | 77.2877(1) | 0.7(1) | | |
| 3.55(5) | 1 | $TbFe_2D_x$ | *C*2/*m* | 247.544(5) | 61.886(1) | 86.5(4) | 3.69 | 3.67 |
| | 2 | $TbFe_2D_x$ | *Fd*-3*m* | 488.80(5) | 61.10(1) | 10.3(2) | 3.38 | |
| | 3 | $TbFe_2D_x$ | *C*2/*m* | 253.00(6) | 63.25(1) | 2.7(1) | 4.35 | |
| | 4 | $Tb_2O_3$ | *Ia*-3 | 1236.34(5) | 77.271(1) | 0.6(1) | | |
| 4.04(5) | 1 | $TbFe_2D_x$ | *C*2/*m* | 252.809(5) | 63.202(1) | 71.5(3) | 4.28 | 4.14 |
| | 2 | $TbFe_2D_x$ | *C*2/*m* | 248.169(6) | 62.042(1) | 24.6(2) | 3.76 | |
| | 3 | $TbFe_2D_x$ | *Fd*-3*m* | 500.67(2) | 62.583(1) | 3.3 (1) | 3.98 | |
| | 4 | $Tb_2O_3$ | *I a* -3 | 1236.39(4) | 77.274(1) | 0.6(1) | | |

A rather good agreement is obtained between the total D content measured by a volumetric method in the samples and that calculated considering the percentage and the D content of each phase obtained using the cell volume expansion (Table S1).